\documentclass[preprint,preprintnumbers,aps,nofootinbib,tightenlines,floatfix,groupedaddress]{revtex4-1}

\usepackage{hyperref}
\usepackage{amsmath,amssymb}
\usepackage{cleveref}
\usepackage{graphicx}
\usepackage{xcolor}
\usepackage[utf8]{inputenc}
\usepackage{bm}
\usepackage{comment}
\usepackage[percent]{overpic}
\usepackage{qcircuit}
\usepackage{csquotes}
\usepackage[section]{placeins}
\usepackage{rotating}
\usepackage{csquotes}
\usepackage{longtable}

\allowdisplaybreaks

\newcommand{\gsim}{\raisebox{-0.7ex}{$\stackrel{\textstyle >}{\sim}$ }}

\newcommand{\Tr}{\ensuremath{\text{Tr}}}
\newcommand{\dif}{\ensuremath{\text{d}}}

\newcount\hour \newcount\hourminute \newcount\minute
\hour=\time \divide \hour by 60
\hourminute=\hour \multiply \hourminute by 60
\minute=\time \advance \minute by -\hourminute
\newcommand{\mydate}{\ \today \ - \number\hour :\number\minute}

\begin{document}

\title{Systematically Localizable Operators \\ for Quantum Simulations of Quantum Field Theories}

\author{Natalie Klco}
\email{klcon@uw.edu}
\affiliation{Institute for Nuclear Theory, University of Washington, Seattle, WA 98195-1550, USA}
\author{Martin J.~Savage}
\email{mjs5@uw.edu}
\affiliation{Institute for Nuclear Theory, University of Washington, Seattle, WA 98195-1550, USA}

\date{\mydate}

\preprint{INT-PUB-19-058}

\begin{figure}[!t]
 \vspace{-1.5cm} \leftline{
 	\includegraphics[width=0.2\textwidth]{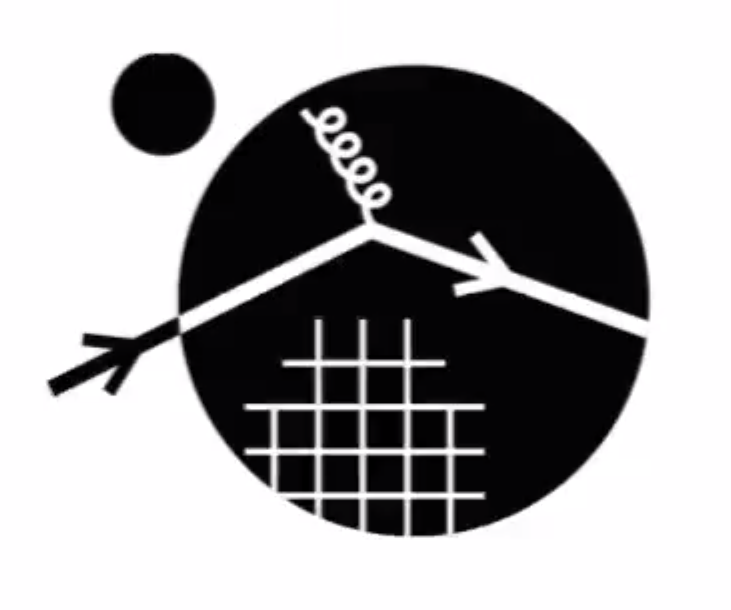}}
\end{figure}

\begin{abstract}
Correlations and measures of entanglement in ground state wavefunctions of
relativistic quantum field theories are spatially localized over length scales set by the
mass of the lightest particle.
We utilize this localization to design digital quantum circuits
for preparing the ground states of
lattice scalar quantum field theories.
Controlled rotations that are exponentially localized in their position-space extent are found to provide exponentially convergent wavefunction fidelity.
These angles scale with the correlation between sites and the classical two-point correlation function, as opposed to the more localized mutual information or the hyper-localized negativity.
We anticipate that further investigations will uncover quantum circuit designs with controlled rotations dictated by the measures of entanglement.  This work is expected to impact quantum simulations of systems of importance to nuclear physics, high-energy physics, and basic energy sciences research.
\end{abstract}
\maketitle

\tableofcontents
\vfill\eject

\section{Introduction}
\noindent
Ubiquitous in modern physics, quantum field theories (QFTs)
are used to
quantitatively
describe physical systems from the dynamics of quarks and gluons in the early universe, to the structure of matter in dense astrophysical objects,
to the structure and properties of common and exotic materials.
The 20th century established the prominence of
renormalizable QFTs with local gauge symmetries in describing the strong and electroweak interactions, three of the four fundamental forces in nature.
Also at this time effective field theories (EFTs) were established as powerful tools to broadly describe
nuclear forces, the properties of hadrons, and lattice regulated field theories over wide kinematic regimes.
While significant progress has been made in
developing precise analytic frameworks and classical numerical calculations of low-lying observables in non-perturbative systems, other classes of observables---notably real-time dynamics and properties of high density systems---remain to be addressed.

Since the early 1980's, it has been imagined that the attributes of quantum systems making their at-scale classical computation so onerous will be more naturally represented on quantum computational devices~\cite{Feynman:1981tf}.
In quantitative support of this perspective, the complexity class BQP (Bounded Error Quantum Polynomial) has recently been shown to extend beyond the regime of classical polynomial-time calculations~\cite{Raz:2019:OSB:3313276.3316315}.
This result indicates that there are problems efficiently accessible to quantum computers but not to classical computers.
While the application of quantum computing to scientific applications is only now beginning and it is anticipated that more than a decade of research and development will be required to perform calculations comparable to results of nuclear physics (NP)~\cite{NPQISrep}
and high-energy physics (HEP) experiments, there is growing excitement at the prospect of addressing real-time dynamics of highly inelastic processes and the structure of high density systems using quantum computers.
Throughout the development toward full-scale quantum simulation,
significant physical insight and advances in classical algorithms are expected to emerge that will impact the NP, HEP, and Basic Energy Science (BES) research programs.

Scalar and pseudoscalar fields play key roles in NP, HEP, and BES at both phenomenological and fundamental levels.
Perhaps the most famous scalar particle is the  $J^\pi=0^+$ isosinglet $\sigma$-field
(with vacuum quantum numbers),
which has been the subject of decades of debate
about its nature but is now firmly established~\cite{Caprini:2005zr}.
The Higgs boson~\cite{Higgs:1964pj,Higgs:1964ia}
has the same quantum numbers as the $\sigma$, and is the remnant of electroweak symmetry breaking driven by a doublet of complex scalar fields that gives rise to the short-range weak interactions
and the long-range electromagnetic interactions~\cite{Weinberg:1967tq,Glashow:1961tr,Salam:1968rm}.
Scalar fields also play a central role in phenomenological
self-consistent relativistic mean-field theories, providing  attractive interactions in the dynamics of large numbers of nucleons (for example, see Ref.~\cite{Chen:2014sca}).
Finally, the pion (and kaon) fields are identified as pseudo-Goldstone bosons associated
with the spontaneous breaking of the chiral symmetries of quantum chromodynamics (QCD)
(for example, see Ref.~\cite{Weinberg:1978kz}).
They have been known to be central to NP since the very earliest days
and dominate the long-distance behavior of the nucleon-nucleon interaction.
In momentum-space, while not required to describe the very low-energy dynamics of nucleon-nucleon (NN)
scattering~\cite{Kaplan:1998we,Kaplan:1998tg,vanKolck:1998bw,Chen:1999tn}
or low-lying inelastic electroweak processes~\cite{Chen:1999tn}, pions are required to correctly describe processes even at modest energies.
Since NN effective field theory (NNEFT) was formulated starting in the early
1990's~\cite{Weinberg:1990rz,Ordonez:1992xp,Ordonez:1995rz,vanKolck:1994yi,Kaplan:1998we,Kaplan:1998tg},
including both pionless and pionful frameworks,
efforts have been ongoing to develop numerical techniques to predict  low-energy
properties of larger nuclear systems based upon NNEFT systematic
power-counting arguments.
Explicitly sampling over dynamical pion fields, e.g., with a lattice discretization of
spacetime (for example, Refs.~\cite{Muller:1999cp,Lee:2004si,Abe:2007fe,Lee:2008fa,Lee:2016fhn}),
constitutes  a persisting line of such developments.
It is anticipated that quantum simulation of low-energy multi-nuclear systems with dynamical pion fields
will provide a quantum advantage
in computing real-time dynamics of systems of interest to NP research, e.g., Ref.~\cite{Holland:2019zju}.
Therefore, it is timely to develop efficient quantum algorithms and quantum circuits for state preparation
and subsequent Hamiltonian time evolution of scalar and (pseudo-)scalar quantum fields on a spatial lattice.

In using  quantum computers to predict observables of importance to NP, HEP, and BES research,
preparing the initial wave function on the quantum register presents one of the major
challenges~\cite{2008arXiv0801.0342K,Jordan:2011ne,Jordan:2011ci,Jordan:2014tma,Jordan:2017lea,Preskill2018quantumcomputingin}.
For example, to determine S-matrix elements of an interacting lattice field theory using the prescription of
Jordan, Lee and Preskill (JLP)~\cite{Jordan:2011ne,Jordan:2011ci,Jordan:2014tma,Jordan:2017lea}, localized wave packets
of the non-interacting theory are initially prepared.
The system is then adiabatically flowed to arrive at localized wave packets in the fully interacting theory, evolved in time through a Trotterized evolution operator, and measured.  The final distribution of particles determines the of S-matrix elements of interest.
While a perfect initialization would ensure that vacuum fluctuations are accurately captured, an imperfect initialization may allow a background of particles to emerge and evaporate during this process.
For example, in calculating the inelastic scattering of two nucleons directly from QCD, a high fidelity preparation of the quantum vacuum is crucial for isolating the process of interest from a complex array of strong interaction backgrounds such as glueballs, pions, and baryon-anti-baryon systems.

Even in the absence of wave packets,
preparation of
the ground state of the non-interacting lattice field theory remains nontrivial.
Focusing on free lattice scalar field theory,
the ground state of this coupled system can be defined by $N$ harmonic oscillators, where $N$
is the number of spatial sites of the underlying lattice that discretizes space.
Such systems of coupled harmonic oscillators have been considered previously from the view point of quantum information science, including the behavior of mutual information and negativity, e.g. Ref.~\cite{Anders:2008a}.
It is most easily represented as the tensor product of the ground state wavefunctions of
each of the momentum eigenstates, which can be Fourier transformed into position space.
Correspondingly, this wavefunction is the sum over products  of oscillator states, one at each spatial site,
coupled through the gradient operator in the Hamiltonian.
This leads to a high-dimensional Gaussian wave function in the continuum limit,
with a non-diagonal covariance matrix in field space.
As the
classical and quantum correlations---for instance the two-point function, its inverse and measures of entanglement---fall exponentially with spatial separation,
the off-diagonal elements in this covariance matrix fall correspondingly.
It is this scaling with separation that allows for a quantum computer to be initialized into the ground state
of a lattice field theory
with the number of operations scaling as a polynomial in the spatial volume.

In this work, we design localizable quantum circuits to prepare the ground states of lattice scalar field theories, both free and interacting,
on digital quantum computers.
Building upon the foundational papers of JLP~\cite{Jordan:2011ne,Jordan:2011ci,Jordan:2014tma,Jordan:2017lea},
Somma~\cite{Somma:2016:QSO:3179430.3179434},
and more recent works~\cite{Klco:2018kyo,Klco:2018zqz,PhysRevLett.121.110504,Klco:2019xro,PhysRevA.99.032306},
the required controlled operations in these circuits
are organized to correspond to spatial separations between field operators---providing
a useful relation between the angles of the controlled rotations and the
classical and quantum correlations of the QFT.
Before presenting numerical results in interacting $\lambda\phi^4$ scalar field theory,
much of our paper explores free field theory because a number of helpful analytic results are found.
Our results are connected to
the growing literature exploring the entanglement in fundamental particle interactions and QFTs~\cite{Calabrese:2004eu,Reznik:2003mnx,Buividovich:2008yv,Donnelly:2011hn,Casini:2013rba,Radicevic:2014kqa,Ghosh2015,Soni:2015yga,Kharzeev:2017qzs,Cervera-Lierta:2017tdt,Witten:2018lha,Beane:2018oxh,Beane:2019loz,Tu:2019ouv}
as well as the literature on area-law entanglement scalings and tensor network
simulatability~\cite{PhysRevLett.91.147902,Hastings:2007iok,PhysRevB.76.035114,Schuch:2008zza,Brandao:2014ppa}.
It is also allows for truncations based upon spatial localization to be systematically implemented at the circuit level,
and for the subsequent loss of fidelity in the prepared wave function to be systematically quantified.

\section{Lattice Scalar Field Theory and Its Correlations}
\noindent
The renormalizable Hamiltonian density describing the dynamics of an interacting real scalar field of mass $m$ is,
\begin{eqnarray}
{\cal H} & = & {1\over 2} \Pi^2 + {1\over 2} ({\bm\nabla}\phi)^2 + {1\over 2} m^2 \phi^2 + {\lambda\over 4!} \phi^4
\ \ \ .
\label{eq:Hlamphi0}
\end{eqnarray}
where the conjugate momentum operator, $\Pi({\bf x})$ has the standard equal-time commutation
relation with the field operator,
$\left[\ \phi({\bf x}) , \Pi({\bf y})\ \right] = i\delta^3 ({\bf x}-{\bf y})$.
For our purposes, we begin by considering  the case of the non-interacting field theory with $\lambda=0$.
In order to numerically evaluate observables in this field theory, space is discretized onto a
cubic grid with a distance between adjacent lattice sites on the
Cartesian axes of $a$ (the lattice spacing) and extent $L$ in each direction.
The number of sites in each spatial direction is $N=L/a$.
In terms of dimensionless quantities and replacing the ${\bm\nabla}\phi$ operator with a
nearest-site finite-difference approximation,
the Hamiltonian density for a d-dimensional non-interacting lattice scalar field theory can be written as,
\begin{eqnarray}
{\cal H} & \rightarrow &
{1\over 2} \ \sum_{\bf j}
\left[\
\hat \Pi^2({\bf j})
+  \hat m^2 \hat \phi^2 ({\bf j})
+
\sum_{\bm\mu}
\left( \hat\phi  ({\bf j}) - \hat\phi  ({\bf j}+ \hat {\bm\mu}) \right)^2
\ \right]
\ \ \ ,
\label{eq:Hlamphi1}
\end{eqnarray}
where $\hat{\bm\mu}$ are unit vectors in the d-spatial dimensions,
and the sum over ${\bf j}$ corresponds to the sum over all of the lattice sites.
We have introduced $\hat m=am$ and $\hat \phi$ to denote the dimensionless mass and
field operator (using powers of the lattice spacing).
Throughout the rest of this work, we will use $\phi$ to denote $\hat \phi$ for simplicity in notation.
Mapping onto a quantum register requires digitizing $\phi$ at each lattice site.
Before addressing such a system it
is interesting to consider the ground state of this lattice system with continuous $\phi$'s.
In the case of periodic boundary conditions (PBCs),   $\phi(i,j,k,L,l,m,n,...)=\phi(i,j,k,0,l,m,n,...)$, where the latin arguments represent spatial variables for the scalar field in arbitrary dimensions that become wrapped periodically in each individual dimension.
When working in the field basis defined by the eigenstates of the field operator $\phi$ at each spatial site,
the conjugate momentum operator can be replaced with the derivative operator in field space,
$\hat\Pi \psi(\phi) \rightarrow  -i {\partial\over\partial\phi}\psi(\phi) $.

As the continuum limit is approached, $a\rightarrow 0$, it is helpful to smear operators over
physically relevant lengths scales, and not restrict them to be localized around the lattice spacing~\cite{Davoudi:2012ya}.
This smearing will make only perturbatively small changes (suppressed by the extent of the smearing) to the value
of observables with support in the infrared,
but will modify the calculated value of observables at the scale of the smearing or higher.
As such,  ${\bm\nabla}\phi$ could be replaced by an operator smeared over a range of lattice sites, such as
\begin{eqnarray}
{\bm\nabla}\phi
& \rightarrow &
w_1 ( \phi  ({\bf j}) - \phi  ({\bf j}+{\bm\mu}))
\ +\
w_2 \ {1\over 2}( \phi  ({\bf j}-{\bm\mu}) - \phi  ({\bf j}+{\bm\mu}))
\ +\ ...
\ \ \ ,
\label{eq:gradsmeared}
\end{eqnarray}
where values of the coefficients $w_{1,2, ...}$ can be chosen to optimize aspects of the subsequent computations.

\subsection{Continuous Lattice Scalar Field in One Spatial Dimension}
\noindent
The issues we are addressing in this work can be demonstrated clearly in 1-dimension,
and  straightforwardly extended to higher dimensions.
The lattice Hamiltonian for continuous-field wavefunctions can be written as,
\begin{eqnarray}
{\cal H} & \rightarrow &
\sum_j^N\
-{1\over 2} {\partial^2\over \partial\phi(j) ^2}
+ {1\over 2} \hat m^2 \phi^2 (j)
+ {1\over 2} \left( \phi  (j) - \phi  (j+1) \right)^2
\ \ \ ,
\label{eq:Hlamphi2c}
\end{eqnarray}
where the nearest-site finite-difference operator is used to represent ${\bm\nabla}\phi$.
The analysis of this lattice system, and those defined in higher dimensions, is well known and can be found in standard texts
on lattice field theory.
Using PBCs, the mass matrix for this system with more than 2 sites has the form of a circulant matrix with first row,
\begin{equation}
  \left(M^2 \right)_1 = \begin{pmatrix}
    \hat m^2+2 &,  -1 &,  0 & \cdots & \cdots & \cdots & \cdots &,  -1\\
  \end{pmatrix}
  \ \ \ ,
\label{eq:MMcirc}
\end{equation}
and the $N$ eigenstates of ${\cal H} $ are spatial plane waves with momentum,
$p_i=0, \pm{2 \pi\over N} , \pm {4\pi\over N} , ... ,+\pi $,
with site dependence
$v_{p_i} (j) = {1\over\sqrt{N}} e^{ i p_i j}$.
The energy-momentum relation resulting from ${\cal H} $
for these momenta is ${\rm E}_i^2 = \hat m^2 + 4 \sin^2 \left({ p_i \over 2}\right) = \hat{m}^2 +\hat{p}_i^2$, defining the hatted momentum variable as that of periodic lattice momentum with ${\rm E}_i$ in lattice units.
The first and last squared energy eigenvalues (at the edges of the Brillouin zone) are singly degenerate,
while the others are doubly degenerate and correspond to states with $+ve$ and $-ve$ momenta.
The vector of fields ${\bm\phi} = (\phi(0), \phi(1), ... ,\phi(N-1))^T$, is related to the (momentum) eigenstates of the $M^2$ matrix via
${\bm\chi}=V {\bm\phi} $, where $V$ is the matrix of the $v_{p_i} (j)$,
with $V.M^2.V^\dagger = {\rm diag}({\rm E}_i^2) = {\bf E}^2$.
The ground state of the coupled system can be written as a tensor product of harmonic oscillators
with energy $E_i$, each in their ground state,
\begin{eqnarray}
| \psi \rangle_0 & = &
| \psi_{\chi_0} \rangle_0 \otimes
| \psi_{\chi_1} \rangle_0 \otimes
\cdots  \otimes
| \psi_{\chi_{N-1}} \rangle_0
\ \ \ ,
\label{eq:unco}
\end{eqnarray}
where
\begin{eqnarray}
\langle \chi_0, \chi_1, .. \chi_{N-1}  | \psi \rangle_0
& = & { {\rm det} {\bf E}^{1/4} \over \pi^{N/4}}\ e^{- {1\over 2} {\bm\chi}^T  {\rm \bf  E} {\bm \chi}}
\ \ \ .
\label{eq:spunco}
\end{eqnarray}
In terms of the eigenstates of $\phi (j)$, this ground state wavefunction of the lattice fields has the form
\begin{eqnarray}
\langle \phi_0, \phi_1, .. \phi_{N-1}  | \psi \rangle_0
& = &
{ {\rm det} {\bf K}^{1/4} \over \pi^{N/4}}\ e^{- {1\over 2} {\bm\phi}^T {\bf K} {\bm \phi}}
\ \ \ ,
\label{eq:spuncophi}
\end{eqnarray}
where ${\bf K}=V^\dagger {\rm \bf  E} V$, which, in general,  has non-zero off-diagonal elements.

\subsubsection{Two-Point Correlation Functions and Lattice Artifacts}
\begin{figure}[!ht]
	\centering
	\includegraphics[width=0.5\columnwidth]{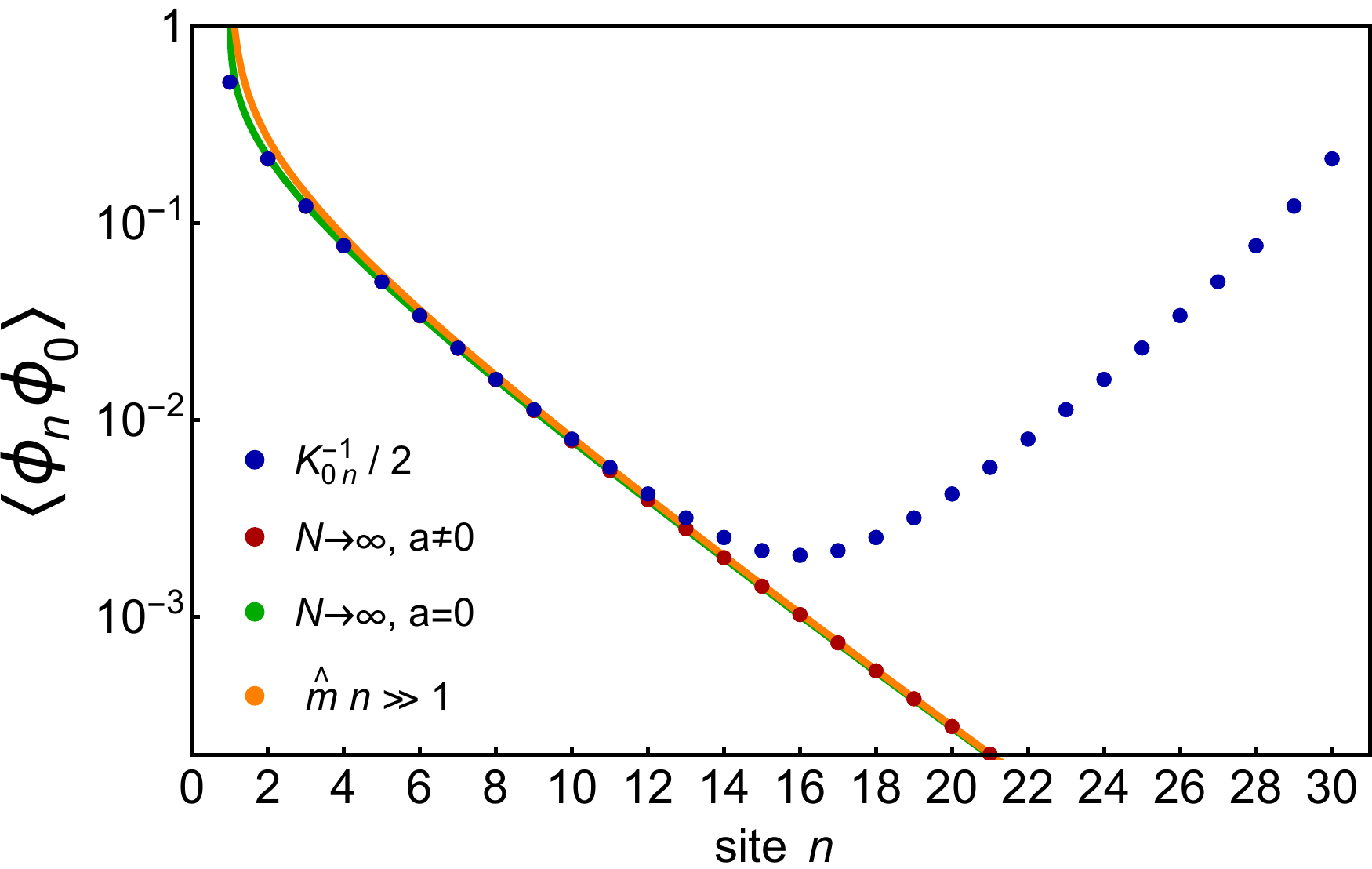}
	\caption{
	The ground state expectation value $_0\langle \psi  | \ \phi_n \ \phi_0\  | \psi \rangle_0$
	for a non-interacting lattice scalar field theory with
	$\hat m=0.3$ calculated on $N=30$ spatial sites (dark blue points).
	Also shown are the infinite volume limit with non-zero lattice spacing  (dark red points),
	the infinite volume and continuum limit (dark green curve), and its asymptotic dependence
	for $\hat m n \gg 1$ (orange curve).
		}
		\label{fig:phiphi}
\end{figure}

\noindent
As a reminder of the impact of finite-volume and finite-lattice-spacing artifacts,
we provide the example of the two-point function,
$_0\langle \psi  | \ \phi_i \ \phi_j\  | \psi \rangle_0$, in the ground state, i.e. the vacuum
expectation value of  field correlations between different lattice sites.
The expectation value of the product of field operators in the ground state of the scalar field theory
can be related to the ${\bf K}$ matrix
defined in Eq.~(\ref{eq:spuncophi}) via
\begin{eqnarray}
_0\langle \psi  | \ \phi_\ell \ \phi_j\  | \psi \rangle_0
& = &
{1\over 2} {\rm \bf K}^{-1}_{\ell j}
\ =\
\sum_\alpha\
{ V_{\ell \alpha }  V_{j \alpha}^*  \over 2 E_\alpha}
\ =\
{1\over N}\
\sum_{p = -(1-{2\over N}) \pi}^{+\pi}\
{e^{i p (\ell-j) }\over 2 \sqrt{\hat m^2 + \hat p^2}}
\nonumber\\
& \rightarrow &
\int_{-\pi}^{+\pi} {dp\over 2\pi}\ {e^{i p (\ell-j)} \over 2 \sqrt{ \hat m^2+\hat p^2}}
\qquad\qquad\qquad
(N=\infty\ \ ,\ \ a\ne 0)
\nonumber\\
& \rightarrow &
\ {1\over 2 \pi} K_0(\hat m |\ell-j|)
\qquad\qquad\qquad
(N = \infty\ \ ,\ \ a = 0)
\nonumber\\
& \rightarrow &
\sqrt{1\over 8\pi}\ {e^{-\hat m |\ell-j|}\over \sqrt{ \hat m |\ell-j| }}
\qquad\qquad\qquad
(N = \infty\ \ ,\ \ a = 0\ ,\ |\ell-j| \rightarrow \infty)
\ ,
\label{eq:phiiphij}
\end{eqnarray}
where the last two lines are the well-known results in the infinite volume and spatial continuum limits and  $K_0(x)$ is a modified Bessel function of the second kind.
In  numerical computations,  finite volume artifacts (which can be represented as sums over periodic images)
and finite lattice-spacing artifacts (which correspond to modifications to the ultra-violet behavior of the theory encapsulated in the Symanzik action~\cite{SYMANZIK1983187}) will be present.
Figure~\ref{fig:phiphi} shows this progression of limits for a non-interacting lattice scalar field theory with
$\hat m=0.3$ calculated on a lattice with $N=30$ spatial sites.
The parameters used in these calculations are chosen so that the scalar particle is contained in the lattice volume,
$\hat m N =9 \gg 1$,
and that it is light compared with the UV cut off of the theory, $\hat m \ll \pi$.
The impact of the PBCs on  $_0\langle \psi  | \ \phi_n \ \phi_0\  | \psi \rangle_0$
is clear from the symmetry about the midpoint of the lattice.
This figure also shows how  the finite-volume, finite-lattice-spacing and large-separation expressions
converge to the numerical results.
The same methods lead to the
infinite-volume, continuum limit of ${\bf K}_{ij}$,
\begin{eqnarray}
{\rm\bf K}_{ij} & \rightarrow &
-{\hat m\over \pi |i-j|}\ K_1(\hat m |i-j|)
\ \ \ ,
\label{eq:Kij}
\end{eqnarray}
where $K_1(x)$ is a modified Bessel function of the second kind.
Appendix~\ref{app:cont4sites} provides the
4-site lattice scalar field theory in somewhat more detail, and
the extension of this methodology to higher dimensions is well known.~\footnote{
In d-dimensions, the energy eigenvalues are related to the scalar mass and momentum through
\begin{eqnarray}
{\rm E}_i & = & \sqrt{
\hat m^2 + 4 \sum_{\bm \nu}\ \sin^2\left({p_{i,\nu} \over 2}\right)
}
\ \ \ ,
\end{eqnarray}
where $\nu$ runs over the spatial dimensions.
The momentum-space fields are related to the position-space fields via a discrete Fourier transform
\begin{eqnarray}
\chi_{\bf p} & = & {1\over\sqrt{N^d}} \ \sum_{\bf j}\ e^{i {\bf p}\cdot {\bf j}}\ \phi({\bf j})
\ \ \ ,
\end{eqnarray}
where ${\bf p}={2\pi\over N} {\bf n}$, where ${\bf n}$ is an d-plet of integers,
with each element of the vector ranging from
$-N/2+1$ to $+N/2$.
}

\subsubsection{Entanglement Entropy and Mutual Information}
\noindent
The  entanglement entropy of a bipartite quantum system quantifies the correlation
between two subsystems, $1$ and $2$,
described by a density operator $\hat{\boldsymbol{\rho}}_{1,2}$, and can be written as,
\begin{equation}
  S(\hat{\boldsymbol{\rho}}_1) = \Tr_1 \left[ \  \hat{\boldsymbol{\rho}}_1 \log \hat{\boldsymbol{\rho}}_1 \ \right]
  \ \ ,
   \qquad {\rm where} \qquad
   \hat{\boldsymbol{\rho}}_1 = \Tr_2 \left[ \  \hat{\boldsymbol{\rho}}_{1,2} \  \right]
  \ \ \ ,
\end{equation}
where $\Tr_j \left[ \ ... \ \right]$ denotes a partial trace over the states in the Hilbert space of system $j$.  When $\hat{\boldsymbol{\rho}}_{1,2}$ represents a pure quantum state, the entanglement entropy is a valid measure of entanglement.
The mutual information (MI) is useful in
quantifying classical and quantum correlations between two subsystems within a larger pure state, $\hat{\boldsymbol{\rho}}_{1,2,3,4,....}$, in which $\hat{\boldsymbol{\rho}}_{1,2}$ is not required, nor expected, to be pure
\begin{equation}
  I(1:2) = S(\hat{\boldsymbol{\rho}}_1) + S(\hat{\boldsymbol{\rho}}_2) - S(\hat{\boldsymbol{\rho}}_{1,2})
  \ \ ,
   \qquad {\rm where} \qquad
  \hat{\boldsymbol{\rho}}_{1,2} = \Tr_{3,4,...} \left[\  \  \hat{\boldsymbol{\rho}}_{1,2,3,4,....}  \ \right]
  \ \ \ .
\end{equation}
Following closely the pioneering work of Srednicki~\cite{Srednicki:1993im},
the entanglement entropy and the MI can be calculated in the
ground state of a non-interacting lattice scalar field theory.
The reduced density operator associated with the field ${\bm\phi}$ and ${\bm\phi}'$,
for a system defined in Eq.~\eqref{eq:spuncophi},
 is
\begin{equation}
  \hat{\bm\rho} ({\bm\phi},{\bm\phi}')
  \propto \int \dif \bm\phi \dif \bm\phi' \ \int \dif \bar{\bm\phi} \exp \left[-\frac{1}{2} \begin{pmatrix}
    \bm\phi & \bar{\bm\phi}
  \end{pmatrix} \mathbf{K} \begin{pmatrix}
    \bm\phi \\
    \bar{\bm\phi}
  \end{pmatrix} - \frac{1}{2} \begin{pmatrix}
    \bm\phi' & \bar{\bm\phi}
  \end{pmatrix} \mathbf{K}
  \begin{pmatrix}
    \bm\phi' \\ \bar{\bm\phi}
  \end{pmatrix} \right]
  \ |\bm\phi\rangle \langle \bm\phi' |
  \ \ \ ,
  \label{eq:reduced12K}
\end{equation}
where $\bm\phi$ and $\bm\phi'$ have dimensionality equal to the number of lattice sites
in the  subsystem, and $\bar{\bm\phi}$ has dimension equal to the number of remaining
lattice sites that are removed via partial tracing.
The overall normalization of $ \hat{\bm\rho} ({\bm\phi},{\bm\phi}')$ is not shown as it is not required for calculating the entanglement properties of the state (comprised of
determinant factors and $\pi$'s that set the trace equal to unity).
After integrating over the $\bar{\bm\phi}$ lattice sites, the reduced density matrix can be written as,
see Appendix~\ref{app:MI},
\begin{equation}
  \hat{\bm \rho} ({\bm\phi},{\bm\phi}')
  \propto
  \exp \left[-\frac{1}{2} \left( \tilde{\bm\phi}^T\tilde{\bm\phi} + \tilde{\bm\phi}'^T \tilde{\bm\phi}' \right)
  + \tilde{\bm\phi}^T \beta' \tilde{\bm\phi}' \right]
  \ \ \ .
  \label{eq:betaprimeRHO}
\end{equation}
The eigenvalues of $\hat{\bm \rho} ({\bm\phi},{\bm\phi}')$  are characterized by $\beta'_i$,
the eigenvalues of $\beta'$, by identifying
$\hat{\bm\rho} ({\bm\phi},{\bm\phi}')$
as a tensor product of $n$ one-dimensional density matrices
$\hat{\bm\rho}_{i} (\tilde\chi_i, \tilde\chi'_i)$,
where $n$ is the number of sites in the reduced space.
The eigenvalues of $\hat{\bm\rho} ({\bm\phi},{\bm\phi}')$ are  multiplicative
combinations of the towers of eigenvalues for the individual oscillators~\cite{Srednicki:1993im},
\begin{equation}
  \xi_i = \frac{\beta'_i}{1 + \sqrt{1-\beta_i^{'2}}}
  \ \ ,  \qquad
  \lambda_{i,n} = (1-\xi_i)\xi_i^n
  \ \ ,
  \qquad \qquad
  \lambda_{\hat{\bm\rho}_1}(\vec{n}) = \prod_{i} \lambda_{i,n_i}
  \ \ \ ,
  \label{eq:RHO1eigenvals}
\end{equation}
where $\lambda_{i,n_i} $ is the energy eigenvalue of the $n^{\rm th}$ eigenstate of the
$i^{\rm th}$ oscillator characterized by $\beta'_i$.
From these eigenvalues, the entanglement entropy
associated with $\hat{\bm\rho} ({\bm\phi},{\bm\phi}')$
is
\begin{eqnarray}
  S(\hat{\bm \rho } )
  & = &
   - \sum_{\vec{n}} \lambda_{\hat{\bm\rho}} (\vec{n}) \log \lambda_{\hat{\bm\rho}} (\vec{n})
 \ =\  -\sum_i \left[ \log (1-\xi_i) + \frac{\xi_i}{1-\xi_i} \log \xi_i \right]
  \ \ .
  \label{eq:summedS}
\end{eqnarray}
From  general expressions in Eq.~(\ref{eq:RHO1eigenvals}) and Eq.~(\ref{eq:summedS}),
the MI between two lattice sites may be determined from a
combination of  calculations with different partitions of the $\mathbf{K}$ matrix.
Figure~\ref{fig:KMINN} shows the MI of a free lattice scalar field theory
 with  $\hat m=0.3$.

It is found, not surprisingly, that the MI falls exponentially with the separation
between the lattice sites~\cite{Casini:2009sr}.
To explore this relation, we re-arrange $\hat{\bm\rho} ({\bm\phi},{\bm\phi}')$ in
such a way to make the inter-site-dependence
explicit, and examine the behavior of the MI in the limit of large separation.
Retaining only two lattice sites, $0$ and $j$,
it is convenient to write $\hat{\bm\rho} ({\bm\phi},{\bm\phi}')$ in a way that makes
clear its tensor product nature in the limit that the two lattice sites become infinitely separated,
\begin{eqnarray}
\hat \rho_0 ( \phi,\phi') & \propto &
 \exp \left[\  -{1\over 2}
\left(
 {\bm\Phi}_0^T  {\bm\Sigma}_\infty {\bm\Phi}_0
 +
 {\bm\Phi}_j^T  {\bm\Sigma}_\infty {\bm\Phi}_j
 \ +\
2   {\bm\Phi}_0^T  {\bm\Xi}_\infty {\bm\Phi}_j
\right)
  \ \right]
 \nonumber\\
 {\bm\Phi}_i^T & = & \left( \phi(i), \phi(i)'\right)
 \ \ ,\ \
 {\bm \Sigma}_\infty \ =\
 \left(
 \begin{array}{cc}
 \gamma_{11} & -\beta_{11} \\
- \beta_{11} &  \gamma_{11}
 \end{array}
 \right)
 \ \ ,\ \
 {\bm \Xi}_\infty \ =\
 \left(
 \begin{array}{cc}
 \gamma_{12} & -\beta_{12} \\
- \beta_{12} &  \gamma_{12}
 \end{array}
 \right)
 \ \ ,
  \label{eq:twosite}
\end{eqnarray}
where $\gamma_{ab}$ and $\beta_{ab}$ are defined in Appendix~\ref{app:MI}, and
where $\gamma_{12}, \beta_{12}\rightarrow e^{-\hat m j}/j^p$ at large separation.
The perturbative analysis given in that Appendix shows that   the leading contribution to the MI vanishes as
$I(0:j) \rightarrow e^{-2 \hat m j}/j^{p'}$ for large-separations~\cite{Casini:2009sr}.
Notably, the MI falls exponentially faster than the two-point correlation function,
i.e. with an argument of $-2 \hat m j$ as opposed to $- \hat m j$.

\subsubsection{Negativity}
The negativity~\cite{Vidal:2002zz} is a measure of entanglement quantifying purely-quantum correlations.
As its name suggests, the negativity is the sum of all negative eigenvalues of the partially
transposed density matrix and may be expressed in terms of the matrix square root
\begin{equation}
  \mathcal{N}( \hat {\bm \rho})
  = \frac{||\hat{\bm \rho}^\Gamma||_1 -1}{2} = \frac{\Tr \sqrt{\left(\hat{\bm \rho}^\Gamma \right)^2} -1}{2}
  = \frac{\sum |\lambda_i| -1}{2} = -\sum_{\lambda_i<0} \lambda_i
  \ \ ,
\end{equation}
where the $\lambda_i$ are eigenvalues of (Hermitian) $\hat{\bm \rho}^\Gamma$
and the unit trace of $\hat{\bm \rho}^\Gamma$ has been employed.

As a simple example,
in the case of a free lattice scalar field theory defined on only two spatial sites as detailed in Appendix~\ref{app:negaTWO},
the two-site negativity is
\begin{equation}
  \mathcal{N} ( \hat{\boldsymbol{\rho}}) = \frac{|{\bf K}_{12}|}{{\bf K}_{11}-|{\bf K}_{12}|+\alpha}
  \ \ , \qquad \alpha =  \sqrt{\det {\bf K}}
  \ \ \ ,
  \label{eq:negativityN2}
\end{equation}
which vanishes when the two sites decouple, ${\bf K}_{12}=0$.

\begin{figure}[!ht]
	\centering
	\includegraphics[width=0.65\columnwidth]{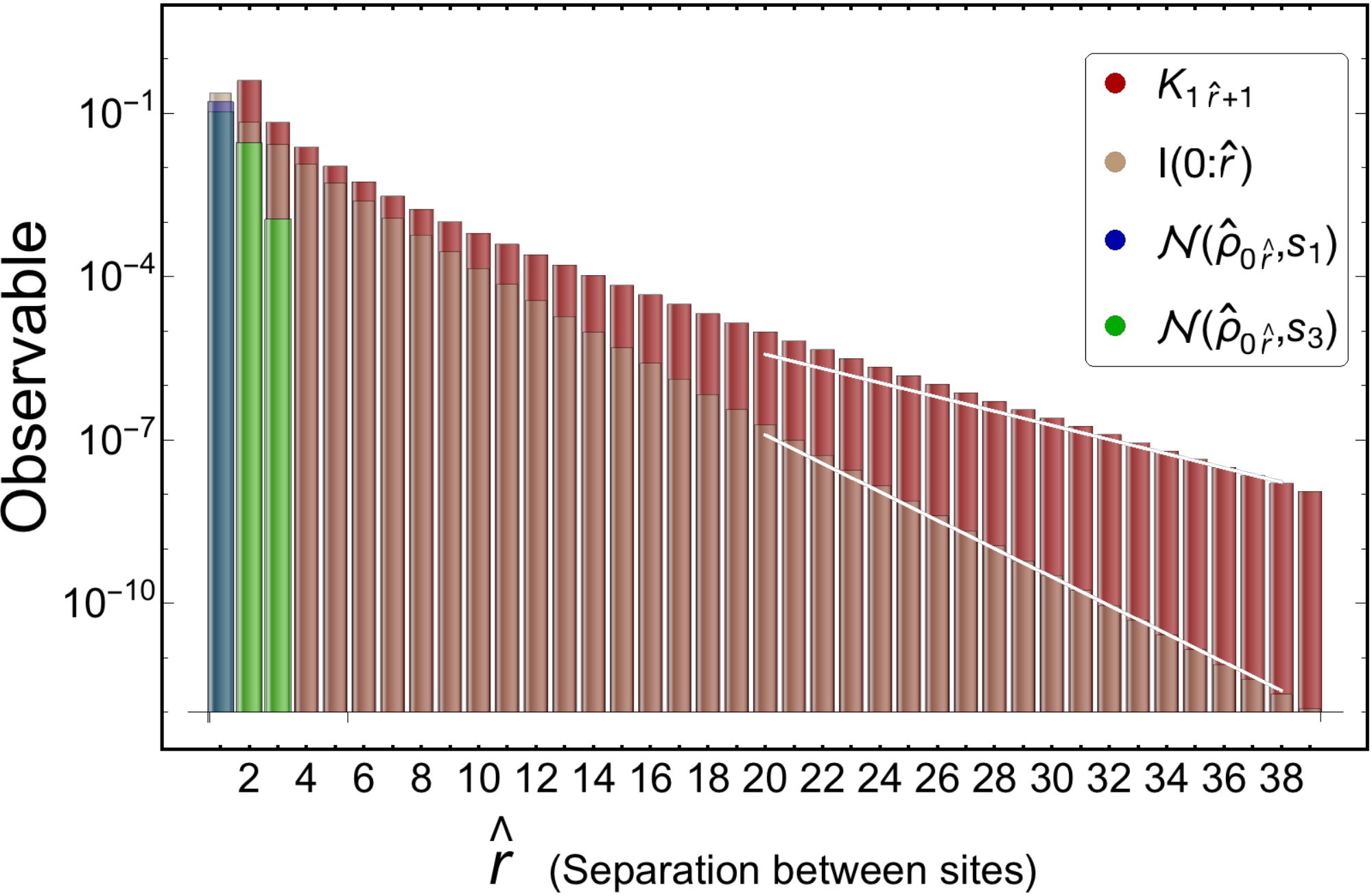}
	\caption{
	The mutual information, $I(0;\hat{r})$, and correlation matrix, ${\rm \bf K}_{1 \hspace{0.1em} \hat{r}+1}$, in the ground state of a free one-dimensional lattice scalar field theory with
	$\hat m=0.3$ and $N=80$ lattice sites with PBCs (gold and red bars, respectively).
	To "guide the eye", the white lines show fits of the form $e^{- \hat m \hat{r}}$ and $e^{-2 \hat m \hat{r}}$.
The blue bar corresponds to the negativity, ${\cal N}(\hat{\boldsymbol{ \rho}}._{0\hat{r}})$, for  ${\bm\nabla}\phi$
	evaluated with a nearest neighbor finite-difference operator, $s_1$.
	The green-bars denote ${\cal N}(\hat{\boldsymbol{\rho}}_{0\hat{r}})$ with ${\bm\nabla}\phi$
	evaluated a difference operator defined on sites separated by up to distance $\hat{r}=3$, $s_3$,
	as defined in Eq.~(\ref{eq:FDops}).
	Both ${\rm \bf K}_{1 \hspace{0.1cm}\hat{r}+1}$ and $I(0;\hat{r})$ have contributions from terms that decay faster than their asymptotic behavior, leading
	to relatively slow approaches to asymptopia.
			}
		\label{fig:KMINN}
\end{figure}
It is interesting to compare $\mathcal{N} (\hat{\boldsymbol{\rho}})$  between the fields at two spatial sites
with the corresponding MI in a large or infinite lattice scalar field theory.
Rather than falling exponentially with spatial separation,
$\mathcal{N} (\hat{\boldsymbol{\rho}}_{0j})$ is localized to extent of the  interaction in the Hamiltonian.
In our present calculations, this scale is set by the extent of the  finite difference gradient operator used to approximate
${{\bm\nabla}\phi}$ in Eq.~(\ref{eq:Hlamphi0}).
The forward-difference gradient operator  shown in Eq.~\eqref{eq:Hlamphi2c},
involving only differences between neighboring sites (denoted by $s_1$),
produces rapidly decaying $\mathcal{N} (\hat{\boldsymbol{\rho}})$,
vanishing  for all  but nearest neighbor negativities.
In this light, $\mathcal{N} (\hat{\boldsymbol{\rho}})$ is sensitive and
characteristic of the ultraviolet properties of the interactions.
This behavior has been observed before in systems of coupled harmonic oscillators~\cite{Anders:2008a}.
It should be said that for small systems, a small non-zero  negativity is found beyond nearest neighbors, but which vanishes rapidly as the extent of the lattices are increased.

To further support these observations, the finite difference operator in Eq.~\eqref{eq:Hlamphi2c}
can be modified or smeared to produce different ultraviolet interactions approximating the gradient.
For example, the symmetric representation with two-site lattice spacing,
${\bm\nabla}\phi \rightarrow \frac{1}{2}\left( \phi  ({\bf x}+1) - \phi  ({\bf x}-1) \right)$,
produces non-zero negativities only at two-site spacings.
In Fig.~\ref{fig:KMINN} we show $\mathcal{N} (\hat{\boldsymbol{\rho}}_{0j})$
calculated from a nearest-neighbor-define ${\bm\nabla}\phi$ operator, $s_1$, and one defined from sites separated by up to three sites
($s_3$),
\begin{eqnarray}
s_1 : && \ ({\bm\nabla}\phi(i))^2 \rightarrow  \left(\phi(i+1)- \phi(i)\right)^2
\nonumber\\
s_3 : &&
\  \left({\bm\nabla}\phi(i)\right)^2 \rightarrow
{1\over 3} \left(\phi(i+1)- \phi(i)\right)^2
\ +\
{1\over 12}  \left(  \phi(i+2)- \phi(i)\right)^2
\ +\
{1\over 27}  \left(  \phi(i+3)- \phi(i)\right)^2
\  .
\label{eq:FDops}
\end{eqnarray}
The negativities associated with $s_{1,3}$ are shown in Fig.~\ref{fig:KMINN} as a function of site separation.
With such localized quantum correlations manipulatable by varying UV completions,
as well as exponentially decaying MI,
it is natural to suspect that a logically-designed quantum circuit for initializing the ground state
of the scalar field will demonstrate corresponding localization of entangling operators.
An explicit digital quantum circuit construction with controlled rotations that are localized on lengths scales set by
 the classical and quantum
correlations that appear in the ${\bf K}$ matrix is the focus of the remainder of this work.

\subsection{Digitized Lattice Scalar Field}
\noindent
Determining properties and dynamics of scalar fields on a quantum computer will require digitizing the scalar field at each lattice site~\cite{Jordan:2011ne,Jordan:2011ci,Jordan:2014tma,Jordan:2017lea}.
There have been a number of recent studies to quantify such digitization artifacts as understood from the
Nyquist-Shannon (NS) Sampling Theorem~\cite{Somma:2016:QSO:3179430.3179434,Klco:2018kyo,Klco:2018zqz,PhysRevLett.121.110504,Klco:2019xro,PhysRevA.99.032306}.
It has been shown that, if the field truncation and digitization are informed by NS saturation and correlated in the computational design as qubits are added, the precision of low energy observables can be made to scale double exponentially with the number of qubits used to represent the field at each lattice site.
We choose to symmetrically digitize the field at each site, using the basis
$\{\  |\phi_i\rangle \ \}$,
defined by eigenvectors of the field operator, $\hat\phi_i$~\cite{Klco:2018zqz,Klco:2019xro}.
For a mapping of each spatial site onto a qubit register of $n_Q$ qubits, and hence a Hilbert space of dimensionality $n_s=2^{n_Q}$,
the Hilbert space of each site in this representation is spanned by $ \left\{ |-\phi_{\rm max} \rangle\ ,\  |-\phi_{\rm max}+\delta_{\phi} \rangle \ ,\
\cdots , \ |-\frac{\delta_{\phi}}{2} \rangle \ ,\ |\frac{\delta_{\phi}}{2} \rangle \ ,\ \cdots,\ |\phi_{\rm max} -\delta_{\phi}\rangle\ ,\ |\phi_{\rm max}\rangle \right\}$ with field-space lattice spacing $\delta_{\phi} = \frac{2 \phi_{\rm max} }{n_s -1}$.

 To recover the digitized version of the analytic entangled  wave function,
 it is convenient to start with a wave function with all elements of the lattice-digitized wave functions appearing with equal weight (e.g. a Hadamard operation has been applied to each qubit on each site),
\begin{eqnarray}
|\psi_i\rangle & = &
{1\over\sqrt{ n_s^{N}}}\
\sum_{i_0,i_1, \cdots ,i_{N-1}=0}^{n_s-1}\
|\phi_0^{(i_0)} \rangle \otimes
|\phi_1^{(i_1)} \rangle \otimes
\cdots \otimes
|\phi_{N-1}^{(i_{N-1})} \rangle
\ .
\label{eq:digiINI}
\end{eqnarray}
We then define a non-unitary operator from Eq.~(\ref{eq:spuncophi}),
\begin{eqnarray}
\hat \Gamma & = & e^{- {1\over 2} \hat {\bm\phi}^T {\bf K}  \hat {\bm \phi}}
\ \ \ ,
\label{eq:spuncophiOP}
\end{eqnarray}
that reproduces the wave function, once appropriately normalized, in the continuum.
The digitized wave function is defined as
\begin{eqnarray}
| \psi_d \rangle & = &
A \hat \Gamma |\psi_i\rangle
\ \ \ ,
\label{eq:spuncophiOPact}
\end{eqnarray}
where $A$ is the normalization constant,
that produces a real wave function across the Hilbert space.
The operator is invariant under the symmetry ${\bm\phi} \rightarrow -{\bm\phi}$, and hence the $n_s^{N}-1$ independent real numbers is reduced to
${1\over 2} n_s^{N} - 1$ independent real numbers due to the reflection symmetry in field space.

\section{Quantum Circuits for Ground State Wavefunctions}
\noindent

The following reminds the reader of a generic quantum circuit for preparing wavefunctions with real amplitudes,
and introduces a systematic restructuring of the operators to naturally embed localized correlations.
The restructuring begins by (at most) doubling the number of quantum operations, in the process making manifest the physically-intuitive removal of long-distance circuit elements.
For the lattice scalar field ground state, where two-point correlation functions decay exponentially with separation,
this localization allows for truncations of the quantum circuit to approximate the ground state with exponentially improvable fidelity.
More generally, this restructuring is expected to be advantageous for position-space qubit representations of field theories exhibiting cluster decomposition or for generic quantum states where classical and/or quantum correlations are suppressed with distance e.g., separation on physical hardware.

\subsection{Controlled Rotations: the $\theta$-Angles}
\noindent
In a previous works~\cite{2008arXiv0801.0342K,Quantumc53:online,Klco:2019xro},
circuits have been established to prepare an arbitrary real, positive wavefunction with a focus on gaussian-distributed amplitudes.
The following circuit is reproduced from Ref.~\cite{Klco:2019xro},
\begin{equation}
  |\psi\rangle = \quad \begin{gathered}
    \scalebox{0.9}{
   \Qcircuit @C=0.2em @R=0.6em {
   |0\rangle \quad &  \gate{R(\theta_0)} & \ctrlco{2} & \ctrlco{2} & \ctrlco{2} &  \qw &   \ctrlco{2} & \qw \\
   &&&&&\hspace{0.2cm}\cdots \\
   |0\rangle \quad & \qw & \gate{R(\vec{\theta}_1)} & \ctrlco{2} & \ctrlco{2} & \qw &\ctrlco{2} & \qw\\
   &&&&&\hspace{0.2cm}\cdots \\
   |0\rangle \quad & \qw &  \qw & \gate{R(\vec{\theta}_2)} & \ctrlco{2} & \qw & \ctrlco{2} & \qw\\
   &&&&&\hspace{0.2cm}\cdots \\
   |0\rangle \quad & \qw &  \qw & \qw & \gate{R(\vec{\theta}_3)} & \qw & \ctrlco{2} & \qw \\
   & \vdots &\vdots&\vdots&\vdots& \stackrel{\text{\begin{rotate}{0}$\ddots$\end{rotate}}}{} \\
   |0\rangle \quad & \qw & \qw & \qw & \qw & \qw & \gate{R(\vec{\theta}_{n_Q-1})} & \qw \\
   }
   }
   \end{gathered}
   \qquad , \qquad
   \begin{gathered}
     \Qcircuit @C=0.2em @R=0.6em {
     R(\theta)
     }
   \end{gathered}
   \quad
   =
   \begin{gathered}
     \begin{pmatrix}
       \cos \theta & -\sin \theta \\
       \sin \theta & \cos \theta
     \end{pmatrix}
   \end{gathered}
   \ \ \ ,
   \label{eq:thetacircuit}
\end{equation}
where the circle-dot controlled operator represents a set of operators each controlled on a different binary state of the control qubits.
This circuit is defined by $2^{n_Q}-1$ rotations and can be implemented with $2^{n_Q}-2$ CNOT gates.
The $\theta_\ell$ vectors at level $\ell$ have length $2^\ell$ with element $\theta_{\ell,k}$
associated with the binary control value of $k$ on the above qubits as read from top to bottom in the qubit register.
The values of the rotation angles are easily defined through ratios of bipartitions of the wavefunction, $|\psi\rangle$,
\begin{eqnarray}
& & |\psi\rangle = \sum\limits_{x=0}^{2^{n_Q}-1} \psi(x)|x\rangle  \hspace{1.8cm} \theta_{\ell, k}  \ = \ \arctan
  \sqrt{
  \frac{
  \sum\limits_{x = x_{\rm min}}^{x = x_{\rm max}}
  \psi(x)^2}{\sum\limits_{y = y_{\rm min}}^{y = y_{\rm max}} \psi(y)^2}}
  \nonumber\\
&&   \begin{array}{ll}
  x_{\rm min}  =    2^{{n_Q}-\ell-1} (2k+1) & \ \ ,\ \   x_{\rm max} =    2^{{n_Q}-\ell } (1+k) -1 \\
  y_{\rm min}  =  k 2^{{n_Q}-\ell} & \ \ ,\ \   y_{\rm max} =    2^{{n_Q}-\ell-1} (2k+1)-1
  \end{array}
  \ \ \ .
  \label{eq:thetas}
\end{eqnarray}
As constructed, the qubit at level-0 is the
\enquote{most-significant} qubit in the binary register.  Its binary value controls whether the amplitude
resides in the first or second half of the wavefunction when represented in a one dimensional
binary-interpreted Hilbert space.  If the wavefunction is symmetric in this computational basis,
Eq.~(\ref{eq:thetas}) indicates that $\theta_0 = \frac{\pi}{4}$.
Analyzing the structure of these rotation angles further, a symmetric wavefunction will be defined by angles satisfying
\begin{equation}
  \theta_{0} = \frac{\pi}{4} \qquad,\qquad \theta_{\ell,j} + \theta_{\ell, 2^{n_Q}-1-j} = \frac{\pi}{2} \ \forall \ \ \ell>0,  j \in \{0,1, ...,2^{\ell-1}-1 \}   \ \ \ ,
  \label{eq:symmetricthetas}
\end{equation}
reducing the rotational degrees of freedom by two.
Returning to the wavefunction from the $\theta$ angles can be accomplished by
%
\begin{equation}
  \psi(x) = \prod_{\ell=0}^{2^{n_Q-1}} \begin{cases}
    \cos \left(\theta_{\ell, \lfloor x 2^{\ell - n_Q} \rfloor} \right) & \text{if } b_\ell = 0 \\[10pt]
    \sin \left(\theta_{\ell, \lfloor x 2^{\ell - n_Q} \rfloor} \right) & \text{if } b_\ell = 1
  \end{cases}
\end{equation}
where $b$ is the binary representation of $x$ read from left to right
e.g., in a two-qubit system, $x = 2$ is associated with $b_0 = 1$ and $b_1 = 0$.
This translation from $\theta_{l,j}$ angles to the wavefunction is a simple result of the $y$-axis rotation gates in Eq.~\eqref{eq:thetacircuit} chosen to parameterize the circuit.

\subsection{Systematically Localizable Operators: the $\alpha$-Angles}
\noindent
If correlations in the target wavefunction are localized,
the structure of the $\theta$-circuit is not optimal in localizing corresponding controlled circuit elements.
To make operator localization explicit,
the $\alpha$ rotation angles are introduced to systematically isolate sensitivity to long-distance controls by decomposing each level, beginning with interactions local with respect to the target qubit.
The basis of this decomposition lies in the circuit identity shown in Eq.~(\ref{eq:circuitIdentity4Alphas}).
\begin{equation}
   \begin{gathered}
   \Qcircuit @C=0.2em @R=1.8em {
   |0\rangle \quad & \ctrlco{1} & \qw \\
   |0\rangle \quad & \ctrlco{1} & \qw \\
   & \vdots \\
   |0\rangle \quad & \ctrlco{1} & \qw \\
   |0\rangle \quad & \gate{R(\vec{\theta}_\ell)} & \qw \\
   }
   \end{gathered}
   = \quad
   \begin{gathered}
   \Qcircuit @C=0.2em @R=1.3em {
   |0\rangle \quad & \qw & \qw &\qw & \qw & \qw & \qw & \ctrlco{1} & \qw \\
   |0\rangle \quad & \qw & \qw &\qw & \qw & \qw & \ctrlco{1} & \ctrlco{1} & \qw \\
   |0\rangle \quad & \qw & \qw &\qw & \qw & \ctrlco{1} & \ctrlco{1} & \ctrlco{1} & \qw \\
   & & &  & & \vdots & \vdots & \vdots \\
   |0\rangle \quad & \qw & \ctrlco{1} & \qw & \qw & \ctrlco{1} & \ctrlco{1} & \ctrlco{1} &\qw \\
   |0\rangle \quad & \gate{R(\alpha_{\ell0})} & \gate{R(\vec{\alpha}_{\ell 1})} & \qw &  \push{ \ \ \cdots \ \ } & \gate{R(\vec{\alpha}_{\ell (\ell-2)})} & \gate{R(\vec{\alpha}_{\ell (\ell-1)} )} & \gate{R(\vec{\alpha}_{\ell \ell} )}  & \qw \\
   }
   \end{gathered} \ \ \ .
   \label{eq:circuitIdentity4Alphas}
\end{equation}
While translating from the $\theta$- to  $\alpha$-circuits increases the number of circuit elements by at
most a factor of two, the modified structure allows the localized correlations and entanglement
of the ground state of the free scalar field to become manifest in the form of exponentially-suppressed
rotation angles (as the associated controls increase in spatial extent).
Thus, systematically truncating operators by locality in the $\alpha$-basis allows such wavefunctions
to be approximated with high fidelity.
The $\alpha$-angles satisfying the above decomposition are:
\begin{equation}
  \alpha_{\ell 0} = \sum_{m = 0}^{2^{\ell}} \frac{\theta_{\ell, m}}{2^{\ell}} \qquad \alpha_{\ell h, k} = \sum_{m = 0}^{2^{\ell-h}-1} \frac{\theta_{\ell, (k + m 2^h) \bmod 2^\ell}}{2^{\ell-h}} - \sum_{m = 0}^{2^{\ell-h+1}-1} \frac{\theta_{\ell, (k+m2^{h-1})\bmod 2^\ell}}{2^{\ell -h+1}}
  \ \ \ .
  \label{eq:alphas}
\end{equation}
The first index $\ell$ indicates the level from which the operator originates and thus the qubit location of the target rotation, the second index $h$ indicates the number of controls on the operator, and the final index $k$ indicates the binary-interpreted value of the associated controls (read from top to bottom).
To determine the $\alpha$'s in terms of $\theta$'s,
one considers the average wavefunction across the range of Hilbert subspaces satisfying the relevant subset of controls.
For the single qubit operator, this is the average rotation of all angles in the level.
For $\vec{\alpha}_{\ell 1}$, this is the average rotation amongst even(odd) binary-interpreted Hilbert subspaces of the previous $\ell$ qubits.
From this mean value, subtractions are made to remove mean value rotations applied previously by operators of higher locality.
The role of the added circuit elements is thus to sequentially isolate the spatially local correlations.
It follows that the relation,
\begin{equation}
  \theta_{\ell k} = \alpha_{\ell 0} + \sum_{h = 1}^{\ell} \alpha_{\ell h, k \bmod 2^h}
  \ \ \ ,
\end{equation}
can be used to translate back to the $\theta$- angles from the $\alpha$-angles.

Note that a decomposition exists also for incomplete collections of localized circuit elements isolated from the fully-controlled operator of Eq.~\eqref{eq:circuitIdentity4Alphas}.
Such a decomposition is implemented by setting $\vec{\alpha}_{\ell h} = 0$ for any set of $h$'s (not including $h = \ell$) and calculating subsequent $\alpha$-angles recursively as
\begin{equation}
  \alpha_{\ell h, k} = \sum_{m = 0}^{2^{\ell-h} -1}  \frac{\theta_{\ell, (k+m 2^h)}}{2^{\ell-h}} - \sum_{m=1}^{h-1} \alpha_{\ell m,k\bmod 2^m} - \alpha_{\ell 0}
  \qquad
  \ \ ,\ \
  \qquad
  \alpha_{\ell 0} = \sum_{k=0}^{2^\ell-1} \frac{\theta_{\ell k}}{2^\ell}
  \ \ \ .
  \label{eq:alphasRec}
\end{equation}
One such partial decomposition useful for the scalar field ground state creates on \emph{site-wise controls}, where the operators controlling on a partial spatial site are set to zero.  For example, in an eight-site lattice with two qubits per site, the $\theta$ operator at level $\ell = 7$ can be, for instance, decomposed into operators with $h$ values of 0, 1, 3, 5, and 7 only.  
An explicit example of the quantum circuit and numerical values of site-wise $\alpha$-angles is provided in Appendix~\ref{app:AlphaExample}.

Without any particular symmetry assumed of the wavefunction,
the $\alpha$-representation naturally exhibits symmetries.
If the full decomposition of all $\vec{\alpha}$'s is implemented
(as opposed to site-wise controls), the following symmetries are present
\begin{equation}
  \alpha_{\ell h, k} = -\alpha_{\ell h, k+2^{h-1}} \ \ \ .
\end{equation}
These angular symmetries maintain the number of independent rotational degrees of freedom at $2^\ell$ for each level
\begin{equation}
  2^0 + \sum_{m = 0}^{\ell-1} 2^{\ell-m-1} = 2^{\ell} \ \ \ .
\end{equation}
Of course, if the level is not fully decomposed with all $\ell+1$ operators, this symmetry
will manifest only for the angles up to the first absent operator.
The $\alpha$-angles present additional symmetries in the case of symmetric wavefunctions
in the binary-interpreted Hilbert space. From Eqs.~\eqref{eq:symmetricthetas}
and~\eqref{eq:alphas}, $\alpha_{\ell 0} = \frac{\pi}{4} \ \forall \ell$.  Additionally,
\begin{equation}
  \alpha_{\ell h,k} = -\alpha_{\ell h,2^h -1 - k}
\end{equation}
leaving only $2^{h-2}$ independent rotational degrees of freedom per localizing operator
and $2^{\ell-1}$ rotations per level
\begin{equation}
  1+\sum_{h = 2}^{\ell} 2^{h-2} = 2^{\ell-1} \ \ \ .
\end{equation}
As expected for a wavefunction with well-defined parity,
and demonstrated intuitively in the replacement of a level
by a single-qubit Hadamard on the most significant qubit (of the binary representation),
the number of angles needed to define the symmetric wavefunction is $2^{n_Q-1}-1$.

It is illuminating to see how the circuit requirements for preparing the GHZ and W states
\begin{equation}
   |\text{GHZ}\rangle_n = \frac{|0\rangle^{\otimes n} + |1\rangle^{\otimes n}}{\sqrt{2}} \qquad ,\qquad |\text{W}\rangle_n = \frac{|10\ldots 0\rangle + |01\ldots0 \rangle + \cdots + |00 \ldots 1 \rangle}{\sqrt{n}}
   \ \ ,
\end{equation}
changes from the $\theta$-representation to the $\alpha$-representation.
In both states, the MI and negativity are translationally invariant.
Recall, however, that the two states are characterized by distinct entanglement patterns and cannot be mixed through local (non-entangling) operations and classical communication (LOCC)~\cite{Greenberger1989,Coffman:1999jd,PhysRevA.62.062314}.
For the $n$-qubit GHZ state, genuine $n$-partite entanglement is present that is fragile to the removal or measurement of any one qubit.
This fragility is quantified by the vanishing of bipartite negativity $\mathcal{N}(\hat{\boldsymbol{\rho}}_1) = 0$ indicating that, after one qubit is removed, all remaining correlations in the GHZ state are classical and quantified by non-vanishing MI.
In contrast, the W-state contains no genuine $n$-partite entanglement though is robust to qubit removal in retaining high degrees of entanglement within reduced density matrices.
As the entanglement structure should dictate the necessary circuit connectivity for state preparation, it is natural to expect these two states to be created by circuits of distinct structure.
Angles capable of initializing the GHZ state are:
\begin{equation}
  |\text{GHZ}\rangle_n : \quad \theta_0 = - \frac{3 \pi}{4} \ \ ,\ \  \theta_{\ell k} = \begin{cases}
    0 & k\bmod 2 = 0 \\
    \pi/2 & k \bmod 2 = 1
  \end{cases}  \rightarrow \quad \alpha_{0} = -\frac{3 \pi}{4} \ \ , \ \  \alpha_{\ell 1,1} = \frac{\pi}{2} \ \ ,
\end{equation}
where each level, $\ell\geq 1$, contains $2^{\ell-1}$ non-zero
$\theta$-rotations of equal importance or a single localized $\alpha$-rotation with $h = 1$.
While it is well known that the GHZ state can be created with such a string of nearest-neighbor controlled operations, deriving this structure with the $\alpha$-angles is encouraging that the proposed transformation allows fragile entanglement to be captured through a series of localized quantum gates.
The angles capable of initializing the W-state are
\begin{equation}
  |W\rangle_n : \qquad \theta_{\ell0} = \arccos \sqrt{\frac{n-\ell-1}{n-\ell}} \ \ \ ,
\end{equation}
where each level contains a single non-zero $\theta$-angle with increasing value as $\ell$ increases and thus as the rotations become highly controlled and non-local.
With a single $\theta$-angle at each level, there is no advantage to using $\alpha$-angles.
As the two-qubit negativity in the W-state, $\mathcal{N}_{W_n}$, is non-zero and constant for increasing qubit separations,
\begin{equation}
  \mathcal{N}_{W_n} = \frac{\sqrt{4 + (n-2)^2} - (n-2)}{2n}  \ \ \ ,
\end{equation}
it is not surprising that fully-controlled rotations are necessary without distance truncation.

\section{Preparing the Non-Interacting Digitized Lattice Scalar Field Ground State}
\label{sec:digitizedprepnumerics}
\noindent

The previous sections have highlighted the potential of preparing the ground state  of a lattice scalar field theory with
quantum circuits constructed in terms of the $\alpha$-rotation angles,
that are designed to correlate with physical spatial separation.
In this section, we elaborate on the attributes and features of these quantum circuits, focusing on the
scaling of the $\alpha_{\ell h,k}$ angles with $h$ the number of controls,
and on the scaling of the  fidelity of the wavefunction as truncations in
 the number of controls and/or the
magnitude of $\alpha_{\ell h,k}$ are imposed.
We numerically examine the forms of the angles, in particular the hierarchies that are present,  and compare with
the $\mathbf{K}$ matrix or two-point correlations, the MI and the negativity.

Before embarking on numerical explorations,
it is worth gaining insight from a perturbative expansion of the
${\bf K}$ matrix, defined in Eq.~(\ref{eq:spuncophiOP}).
While it might have been helpful to establish analytic results without perturbatively expanding in separation, the appearance of
$\arctan( )$ functions in the relation between angles and values of the wavefunction render such results unwieldy.
By assigning a parametric scaling of  ${\rm \bf K}_{ij}\rightarrow Q^{|i-j|} {\rm \bf K}_{ij}$,
where $Q$ is introduced as an expansion parameter
($Q=1$ is set at the end),
and expanding Eq.~(\ref{eq:spuncophiOP}),
the angles $\alpha_{\ell h,k}$ can be written as a perturbative expansion in powers of $Q$.
Using Eq.~(\ref{eq:Kij}), it is straightforward to show that the controlled rotation angles  scale as
$\alpha_{\ell h,k} \sim Q^{\hat r} \sim {\rm \bf K}_{1 \hat r+1},  {\rm \bf K}_{1 2} {\rm\bf K}_{1 \hat r}, ..., {\rm \bf K}_{1 2}^{\hat r}
\rightarrow \hat m K_1( \hat m \hat r )/\hat  r \rightarrow  \sqrt{\hat m}e^{-\hat m \hat r}/\hat r^{3/2}$, where
$\hat r= \lceil (h - \ell \bmod n_Q) /n_Q \rceil$ is the spatial distance (in lattice units) between the lattice sites of the
maximally separated qubits in the controlled operation.
This analysis reveals that the $\alpha$-angles scale in essentially the same way as the $K$ matrix and two-point correlation function,
not with the MI or negativity.

Take, for example, the controlled rotations acting on the highest qubit of the 3rd spatial site
($l=11$, as counting starts from $l=0$) for a lattice scalar field theory with $n_Q=4$ qubits per site.
The spatial separation vanishes, $\hat r=0$, for $h=0,1,2,3$ and hence all of the rotation angles correspond to interactions within a spatial site, scaling as $\alpha_{\ell h,k}\sim {\cal O}(1)$ (determined by ${\rm K}_{11}$).
The $h=4,5,6,7$ rotation angles are controlled by the adjacent site, with $\hat r=1$,
and  scale as $\alpha_{\ell h,k}\sim {\cal O}({\rm K}_{12})$.
The $h=8,9,10,11$ rotation angles are controlled by the site that is two lattice spacings separated, $\hat r=2$,
and scale as  $\alpha_{\ell h,k}\sim {\cal O}({\rm K}_{13})$.
Expanding to higher orders in the $Q$-expansion shows that the angles receive additional contributions that are further suppressed
by powers of ${\rm K_{ij}}$ compared to the leading order contributions.
For a generic lattice system with sufficiently large spatial volume,
$\hat m N \gg 1$ and with control operations between the first and midpoint lattice site scaling as $\sim \sqrt{\hat m} e^{-\hat m N/2}/( N/2)^{3/2}$ ,
this perturbative expansion demonstrates the potential of mappings and
quantum circuits that are arranged to reflect the underlying physical system when working to a pre-determined
level of precision in a computation.

\begin{figure}[!ht]
	\centering \includegraphics[width=0.407\columnwidth]{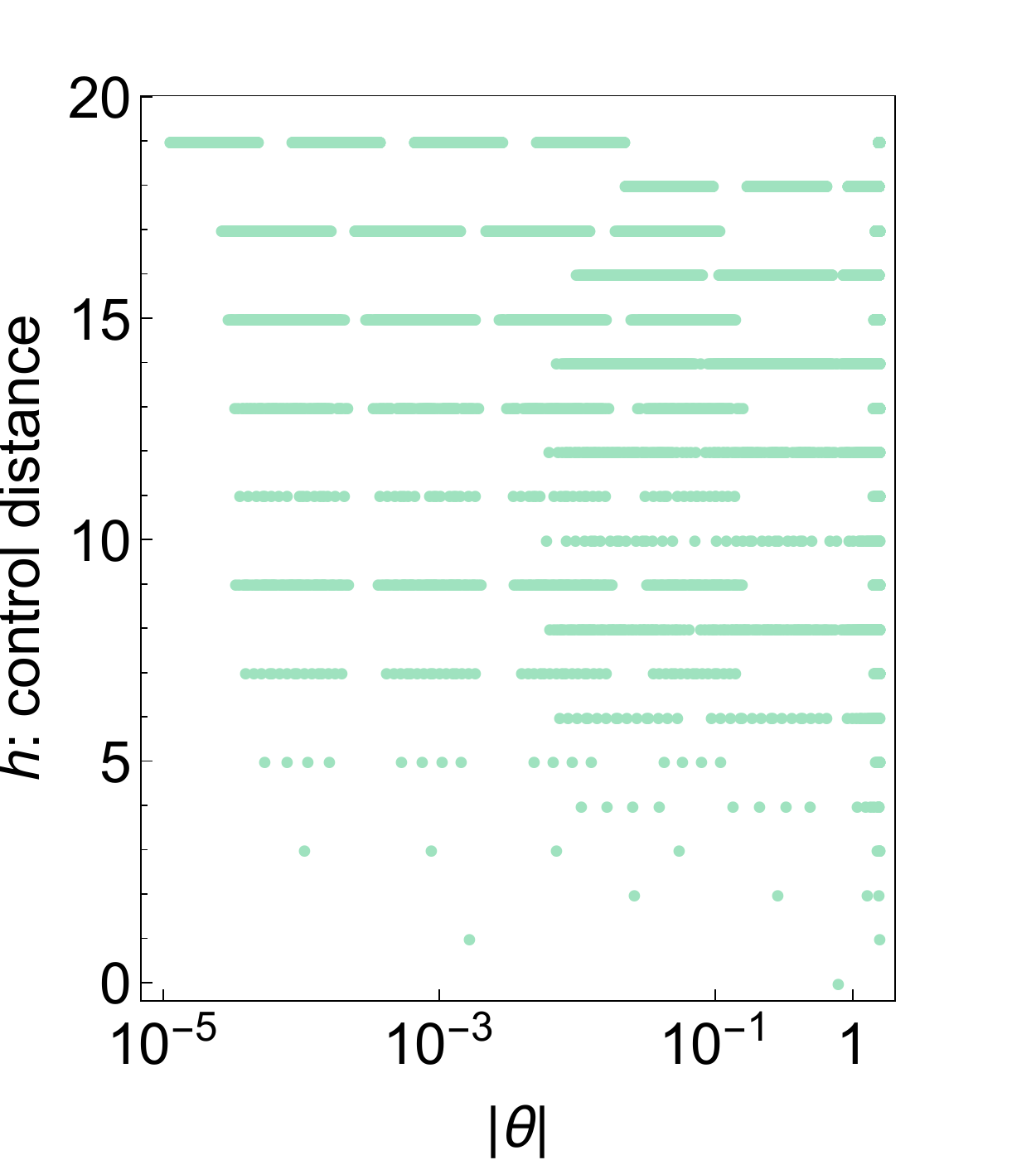}
	 \includegraphics[width=0.4\columnwidth]{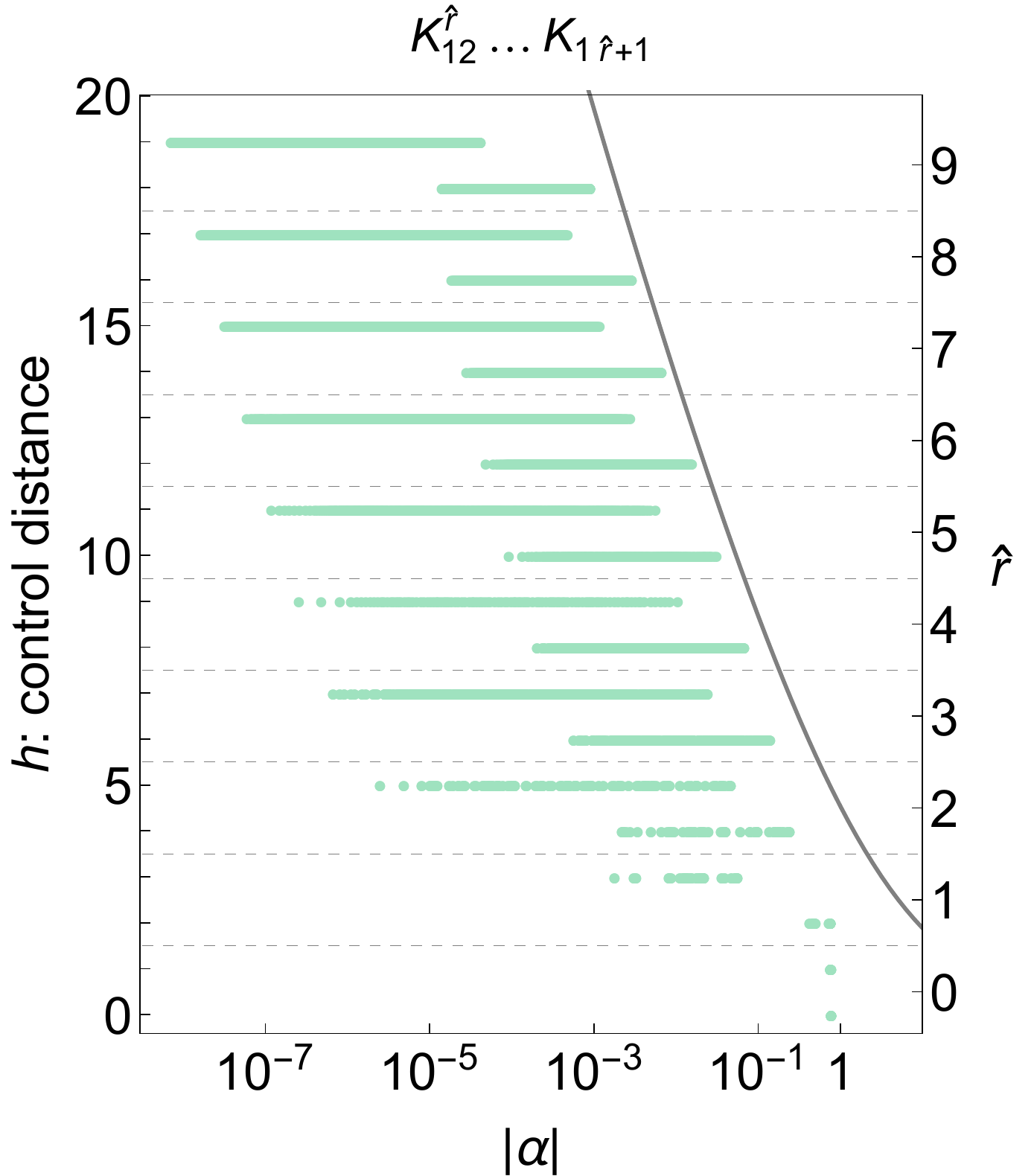}
	\caption{
	The magnitude of the controlled rotation angles $\theta_{\ell,k}$ (left panel) and  $\alpha_{\ell h,k}$ (right panel)
	as related to the control distance, $h$,
	(directly related to spatial separation) for a ten-site ($N=10$) spatial lattice with two qubits per site, $n_Q=2$.
	The light green points show the results of  numerical computations
	while the light-grey curve (right panel) shows the (normalized) theoretical asymptotic form of the inverse two-point function,
	$|\alpha|_{\rm max}(\hat{r}) \sim   M_\alpha  K_1(  M_\alpha \hat r )/ \hat r \rightarrow {\rm \bf K}_{1 \hspace{0.1em} \hat r+1}$ with $M_\alpha$ tending to the scalar mass $\hat{m}$ at large volumes.
		}
		\label{fig:alphahscaling}
\end{figure}
To numerically verify the scaling results obtained  with this perturbative expansion, we analyze a digitized free lattice scalar field theory with open boundary conditions, with
the field at each site encoded onto two qubits, $n_Q=2$, ten spatial sites, $N=10$, a field truncation of $\phi_{\rm max} = 3.5$, and a mass of $\hat m=0.3$ ($\hat{m}N = 3$).
The relation between the $\alpha$-angles and the spatial extent of the associated controls is shown in
the right panel of
Fig.~\ref{fig:alphahscaling}.
For the sake of comparison,
the corresponding relation between the $\theta$-angles and control distance are shown in the left panel of Fig.~\ref{fig:alphahscaling},
which makes clear the absence of the correlations exhibited by the $\alpha$-angles.
The magnitude of  the $\alpha$-angles at a given control distance typically span orders of magnitude, but are bounded above by
$|\alpha|_{\rm max}(\hat{r}) \sim M_\alpha K_1( M_{\alpha} \hat r )/ \hat r \rightarrow {\rm \bf K}_{1 \hat r} $ with  $M_\alpha$ tending to $\hat{m}$ at large volumes, as expected from the perturbative analysis.
For this lattice volume and scalar mass, shown in Fig.~\ref{fig:alphahscaling}, the extracted value is $M_{\alpha} \sim 2 \hat{m}$.
The numerical exploration of this circuit structure at this and larger masses,
consistent with the perturbative analysis, indicates that the angles scale with $\mathbf{K}$ and the inverse two-point correlation function once lattice artifacts become negligible.
Interestingly, this suggests  that there may be an even more efficient quantum circuit design with angles scaling with
the measures of entanglement as opposed to the correlation functions.

\begin{figure}[!ht]
	\centering \includegraphics[width=0.8\columnwidth]{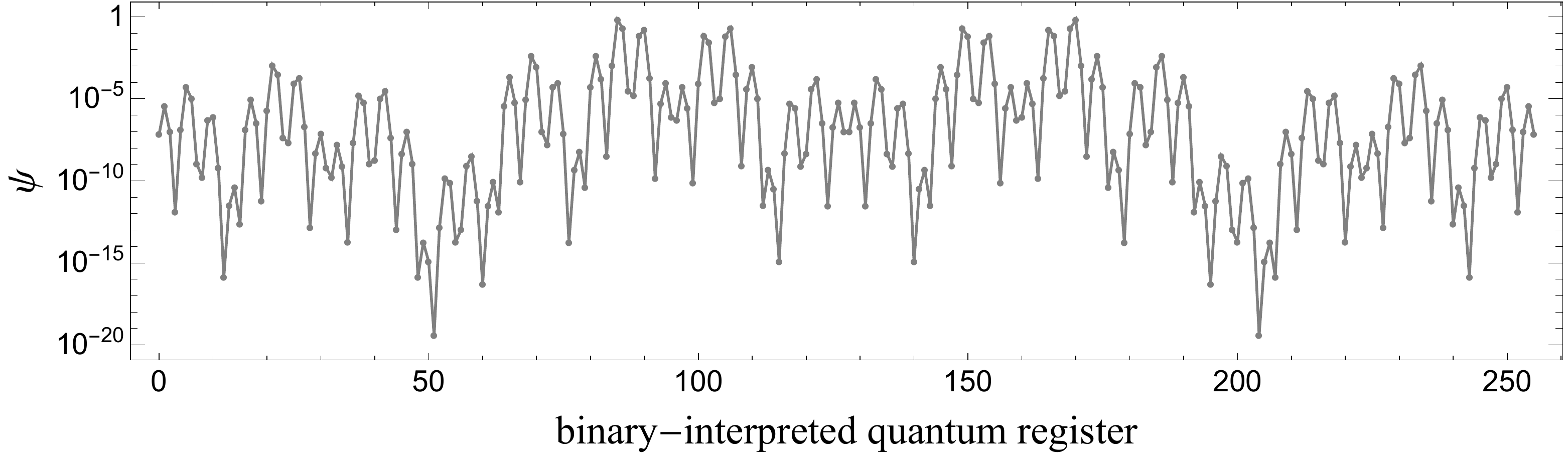}
\includegraphics[width = 0.19\columnwidth]{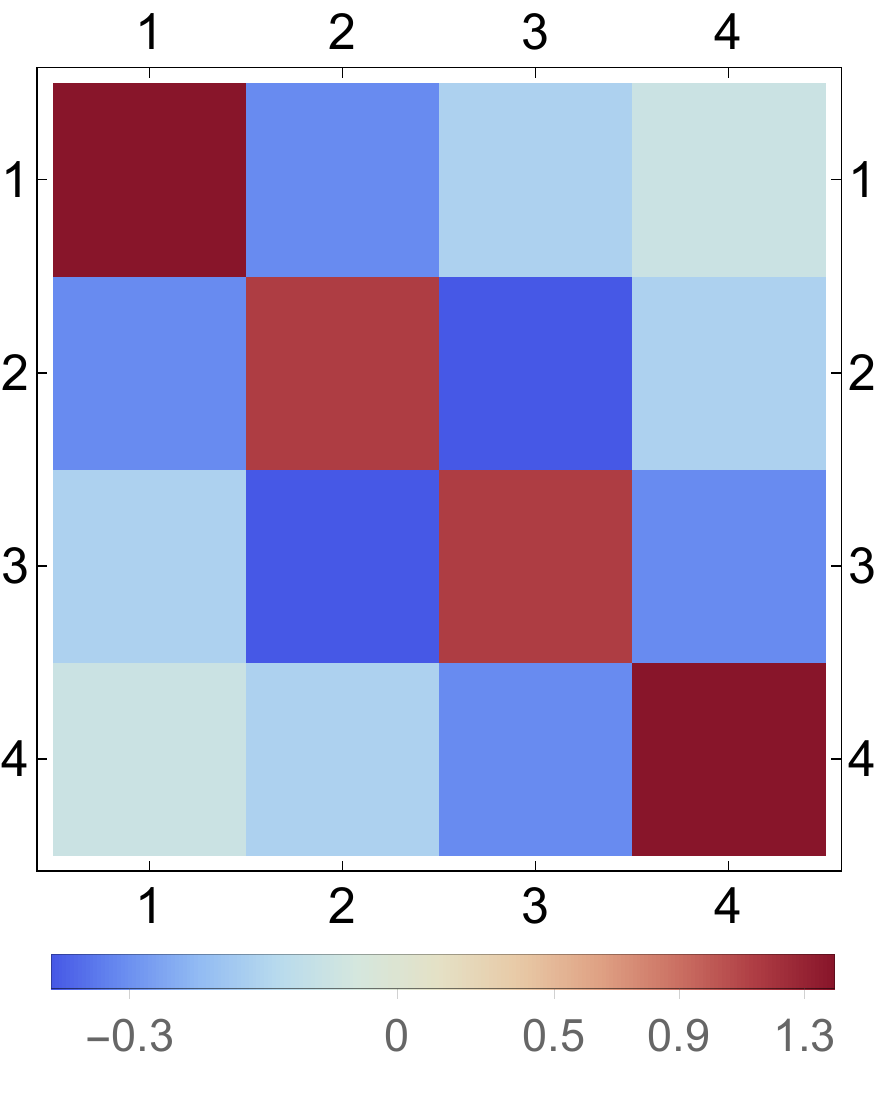}
	\caption{
	The wave function (left) and associated correlation matrix $\mathbf{K}$ (right) of the
	one-dimensional digitized lattice scalar field theory with
	 two qubits per site ($n_Q = 2$), four spatial sites ($N = 4$), $\hat m=0.3$, $\phi_{\rm max}=3.5$,
	 and with open boundary conditions imposed on the field.
		}
		\label{fig:psi4with2}
\end{figure}
While we have shown that the $\alpha$-angle scaling with control distance is consistent with $\mathbf{K}$ and the inverse two-point function,
this relation is only helpful if it provides a wavefunction from the
truncated circuit application
that is perturbatively close to the true wavefunction, and with a quantifiable associated systematic error.
It is straightforward to numerically determine the
fidelity of a wavefunction created with a  quantum circuit
that is truncated in
the number of control elements applied to any given qubit, or in the tolerance of the magnitude of any given rotation angle.
Truncations in each, or both, can be determined in order to perform a computation to a pre-determined level of precision,
that is exponentially convergent.

\begin{figure}[!ht]
  \centering
  \includegraphics[width=0.6\textwidth]{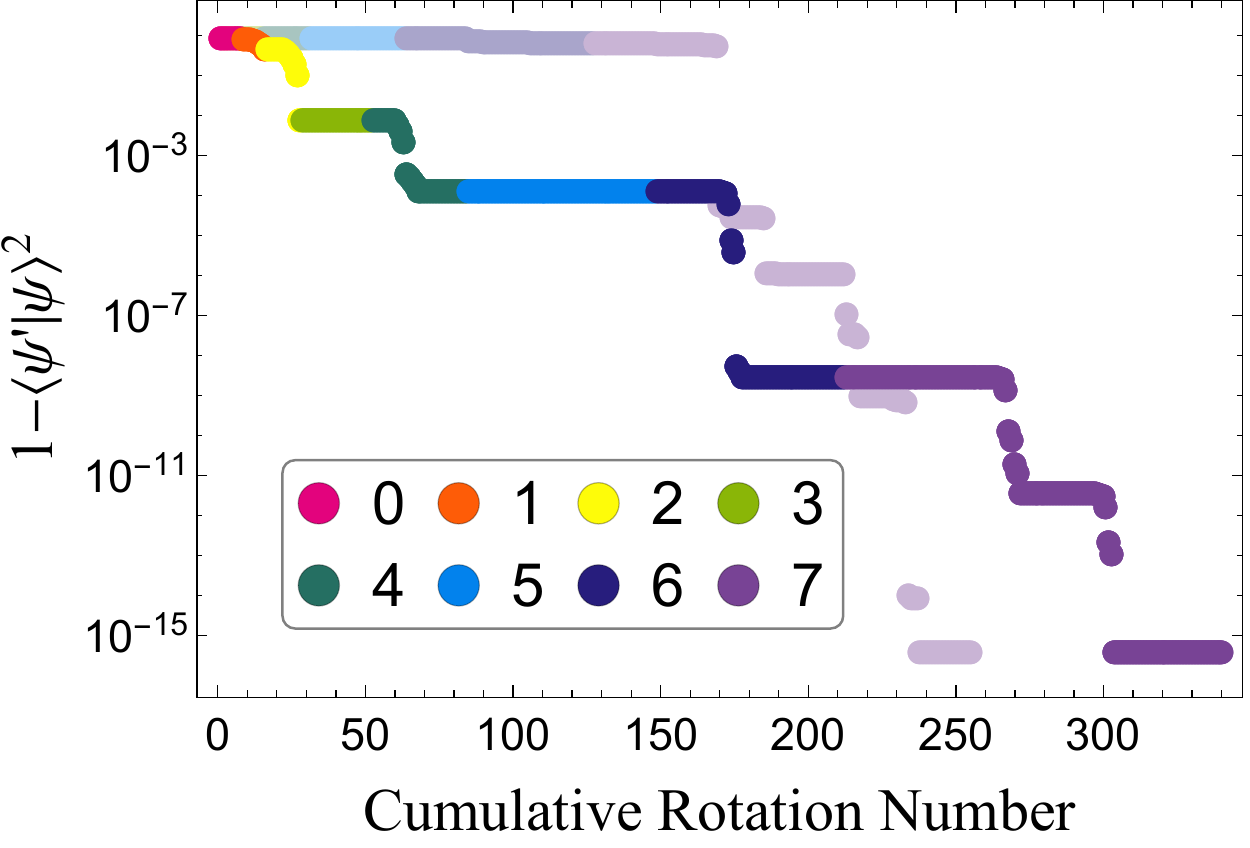}
  \caption{
  Fidelity of the prepared ground state wavefunction as a function of the $h$-organized $\alpha$-angle (dark) and $\theta$-angle (light) truncation value for a lattice scalar theory  with $\hat m=0.3$, $N = 4$, $n_Q = 2$, and $\phi_{\rm max} = 3.5$.
  Controlled rotation angles $\alpha_{\ell h,k}$ are set equal to zero when their magnitude is below the truncation value.
  Points are colored according to, $h$, the number of controls on the associated operator for the rotation angle at the truncation boundary. }
  \label{fig:alphaTruncation_N4nQ2}
\end{figure}
To explore the impact of these truncations, we have studied a system of four spatial sites ($N=4$)
with fields digitized with two qubits per site ($n_Q=2$) and open boundary conditions.
The ground state wave function of this system with $\hat m=0.3$ ($\hat{m}N = 1.2$) and $\phi_{\rm max}=3.5$ is shown in Fig.~\ref{fig:psi4with2}.
This theory has similarities with the two-site ($N=2$) lattice systems with $n_Q=2$ that is detailed in
Appendix~\ref{app:Dig22sites}, whose wavefunction is displayed in Fig.~\ref{fig:psi22}.
Figure~\ref{fig:alphaTruncation_N4nQ2} shows the monotonically decreasing wavefunction fidelity resulting from the $\alpha$-angles and $\theta$-angles ordered first by control distance and then by magnitude.
That is to say that
the $\alpha_{\ell h,k}$ and $\theta_{h,k}$ with $h$-values above and magnitudes below a given truncation value
are set equal to zero.
We observe that, modulo the notable step-like structure, the $\alpha$-circuit allows
the wavefunction to be systematically improved by incorporating circuit elements in an order dictated by the locality and magnitude of their associated rotations.
Therefore, we can limit the number of $\alpha$-angles by defining a hardware connectivity and a threshold below which the angle is set equal to zero,
and there is a direct, but structured, relation between the truncation value in spatial separation and the fidelity of the prepared wavefunction.
Shown in lighter shades of color in Fig.~\ref{fig:alphaTruncation_N4nQ2}, this feature is not present in the $\theta$-angles, where substantial improvement in the fidelity is not achieved until the maximally delocalized rotation operators are included.  For limited connectivity and number of operators, the $\alpha$-circuit is capable of initializing wavefunctions with fidelity improved by multiple orders of magnitude.

\begin{figure}
  \centering
  \includegraphics[width=0.6\textwidth]{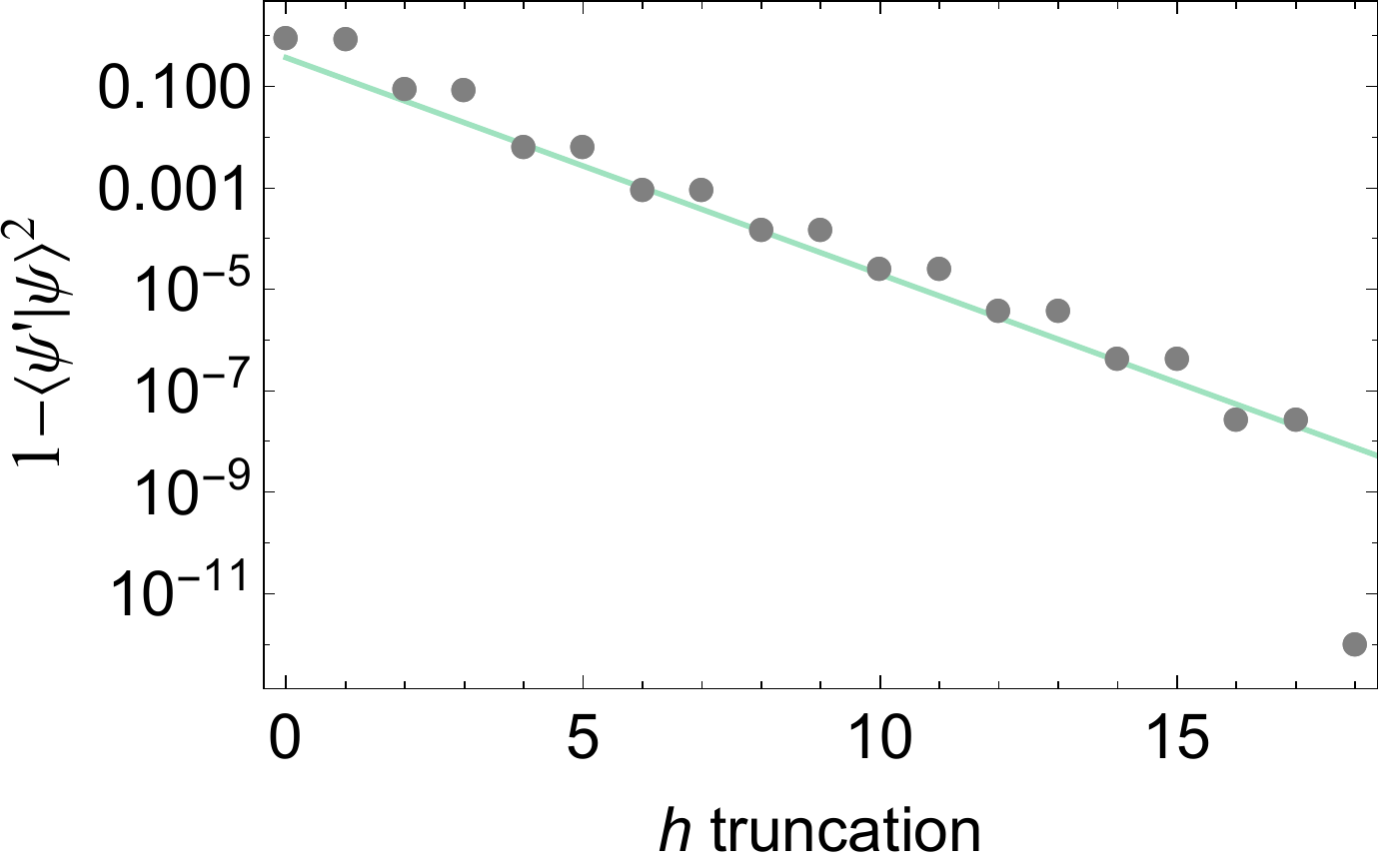}
  \caption{
  Fidelity of the prepared ground state wavefunction as a function of the  control distance truncation
  for a lattice scalar theory
  with $\hat m=0.3$, $N = 10$, $n_Q = 2$, and $\phi_{\rm max} = 3.5$.
  The green line corresponds to a fit with the form $ Z e^{- \eta  (h_{\rm trunc}+1)}$ with $Z = 0.97(10)$ and $\eta = 0.98(2)$ of the even $h_{\rm trunc}$ calculations between $h_{\rm trunc} = 4,14$.
  }
  \label{fig:hTruncation_N4nQ2}
\end{figure}
Figure~\ref{fig:hTruncation_N4nQ2} shows the fidelity of the prepared
wavefunction as a function of control distance truncation, $h_{\rm trunc}$.
A step-like behavior is observed, which is related to the number of qubits defining the field at a single lattice site, $n_Q$.
For these parameters $(N = 10, n_Q = 2, \hat{m} = 0.3, \phi_{\rm max} = 3.5)$, the results show exponential convergence of the wavefunction such that a high fidelity wavefunction can be prepared even with limited spatial connectivity of quantum hardware.

The reorganization of the quantum circuit used to prepare a general ground state of a lattice scalar field theory,
from the $\theta$- to the $\alpha$-parameterization,
allows for truncations in the number of controls or the number of rotation operations, or both,
with a pattern that follows $\mathbf{K}$ and the inverse two-point correlation function.
These truncations can be  removed to systematically improve the fidelity of the wavefunction prepared in the quantum register.
These attributes make  $\alpha$-angle quantum circuits preferable for initializing quantum simulations of scalar field theories.

Our free-field calculations have been performed on relatively
small systems, with only a modest number of (classically-simulated) qubits.
As a result,  field-digitization, lattice artifacts and finite-volume effects are expected to be significant.
Figure~\ref{fig:KMINN} shows the size of finite-volume effects and the separations required for the inverse two-point
function to approach its asymptotic behavior,
which occurs at distances larger than the volumes we have considered in the extraction of the
$\alpha$-angles.
A fit to the mass associated with the exponential fall off of the
control sensitivity in the angles furnishes a mass that differs from than the lightest
mass in the spectrum.  As the discrepancy decreases with increasing input scalar mass, it is expected to be attributed to finite-volume effects.
However, more thorough  investigations
utilizing significantly larger classical computational resources  are required to firmly establish this to be the case.

\section{Interacting $\lambda\phi^4$ Scalar Field Theory}
\label{sec:Lam}
\noindent
The localization method  explored in previous sections using non-interacting lattice scalar field theories
can be implemented for  interacting field theories, such as $\lambda\phi^4$.
While closed form analytic derivations are generally unavailable for interacting field theories,
the applicability of the circuit localization can be demonstrated numerically.

There are a number of available paths forward.
The Hamiltonian for the continuum-field lattice theory can be diagonalized with the necessary
imposition of momentum mode truncations, $\Lambda_k$, building upon the lowest energy basis state given in Eq.~(\ref{eq:unco}).
The non-diagonal Hamiltonian matrix elements are then computed from the interaction term, $\lambda\phi^4$.
The ground state wave function, after transforming back into position space, would then be sampled on the qubit digitized lattice
sites following the method employed above to analyze the non-interacting scalar field theory.
However, for this demonstration it was found to be sufficient to work directly in the digitized space.
The lattice Hamiltonian operator in Eq.(\ref{eq:Hlamphi1}) is supplemented with an interaction term of the form,
\begin{eqnarray}
\hat {\cal H}_{\rm int} & = & {\lambda\over 4 !} \ \sum_{\bf j}\ \hat \phi^4 ({\bf j})
\ \ \ ,
\label{eq:lamphi4int}
\end{eqnarray}
and matrix elements of the full Hamiltonian are computed in the eigen basis of the $\hat\phi$ operator, given in Eq.~(\ref{eq:digiINI}).
As discussed previously~\cite{Jordan:2011ci,Somma:2016:QSO:3179430.3179434,PhysRevLett.121.110504,Klco:2018zqz},
the $\hat\Pi$ operator can be replaced with a discrete operator in the digitized space of each lattice site
without introducing polynomial-scaling digitization errors,
as presented in Appendix~E of Ref.~\cite{Klco:2018zqz}, with either periodic or twisted boundary
conditions in field space.
The Hamiltonian matrix for systems up to nine spatial sites was diagonalized using
Mathematica's intrinsic functions and also a Lanczos algorithm
on a classical computer in the sector containing the ground state and the sector containing a single-particle excitation (at rest).
The initial vectors (interpolating operators) used to start the Lanczos iteration
corresponded to one with equal amplitudes (the ground-state sector),
and to one with an application of $\sum\limits_{\bf j}\hat \phi({\bf j})$ to
the ground-state initial vector.

\begin{figure}[!ht]
	\centering \includegraphics[width=0.4\columnwidth]{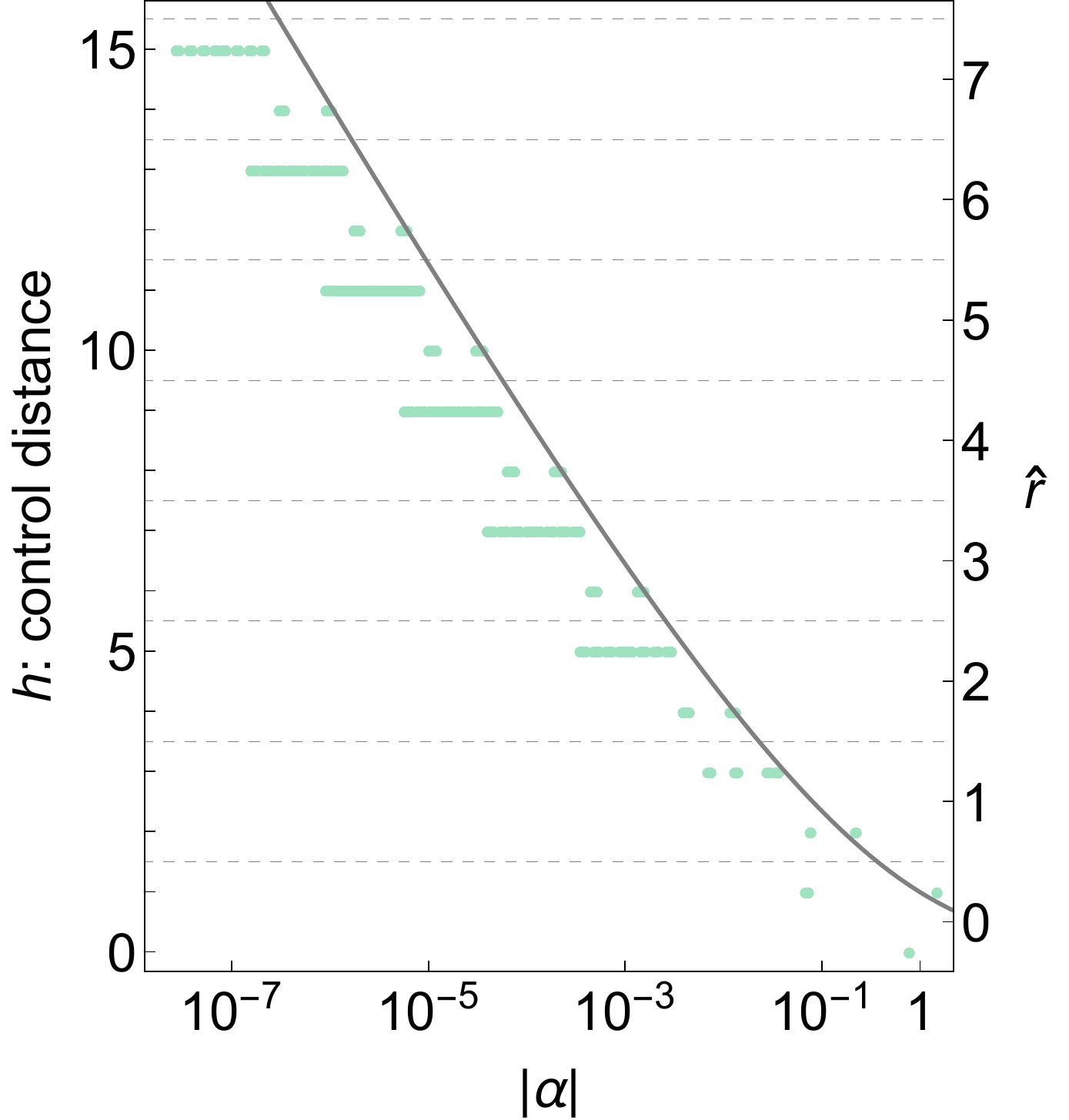}
	\ \ ,\ \
	\includegraphics[width=0.4\columnwidth]{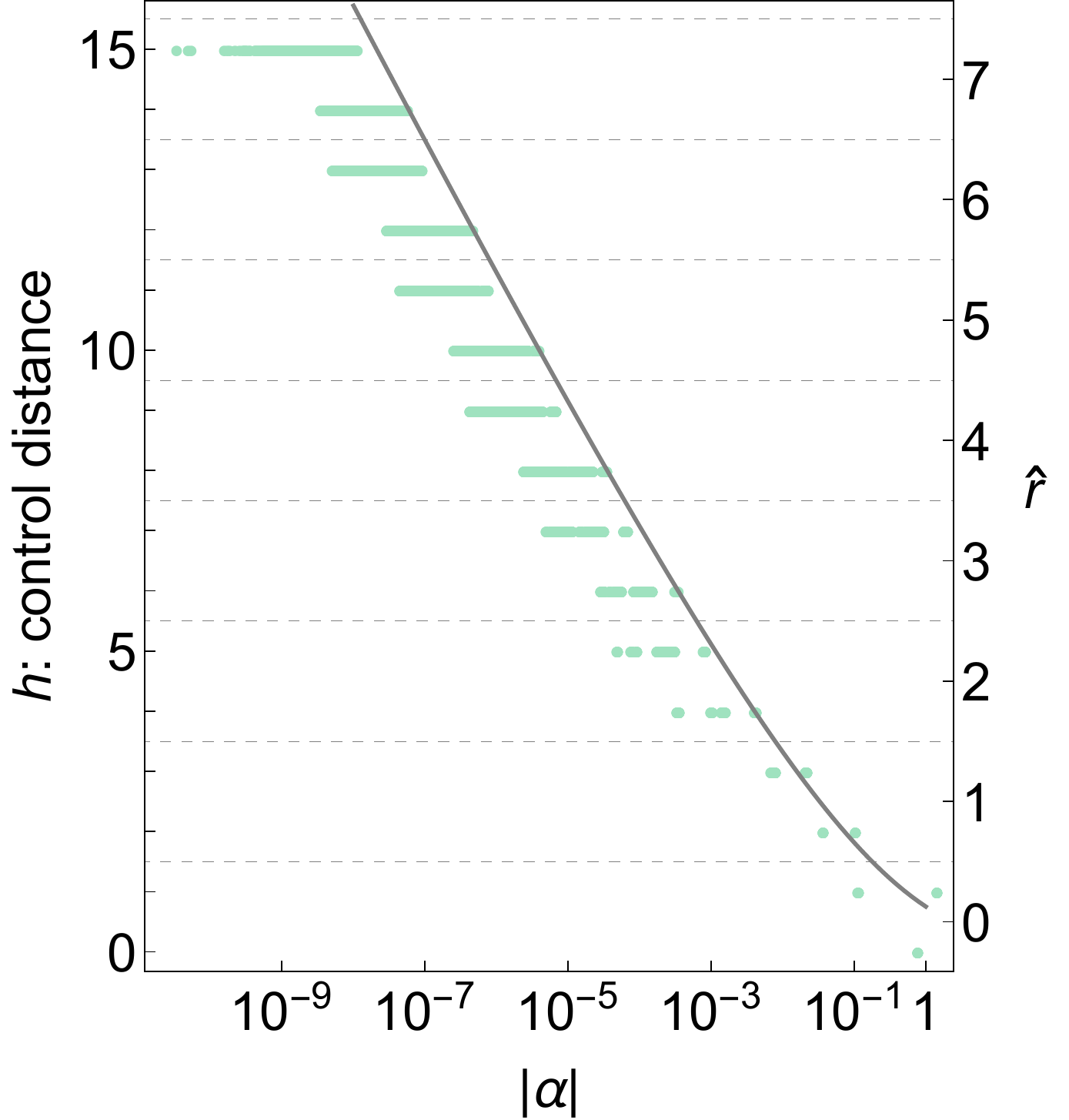}
	\caption{
	$\alpha$-angles as a function of  control distance, $h$.
	The  left panel shows the results
	(green points)
	from free field theory with $N=8$, $m=1.6$, $\phi_{\rm max}=1.7$ and $n_Q=2$,
	while the right panel shows those from interacting $\lambda\phi^4$ field theory
	with the same parameters as the free results except for $\phi_{\rm max}=1.3$ and with $\lambda=32$.		
	The grey curve corresponds to fits
	of the form $c_1 e^{- M_\alpha \hat{r}}/\hat{r}^{3/2}$
	 to the maximum $\alpha$ angle for each $h$.
		}
		\label{fig:alphahscalingLam}
\end{figure}
For demonstration, the $\alpha$-angles are shown in Fig.~\ref{fig:alphahscalingLam} as a function of control distance for
both free field theory and interacting field theory, and with a heavier mass than considered earlier.
We choose $N=8$, $m=1.6$, $\phi_{\rm max}=1.7$ and $n_Q=2$, with $\phi_{\rm max}=1.3$ and $\lambda=32$ in the interacting theory,
after tuning (as discussed in Ref.~\cite{Klco:2018zqz}) using an $n_Q=3$ system on a smaller lattice.
The single-site projected wavefunctions are well localized within the chosen $\phi_{\rm max}$ field truncations.
The results are shown in Fig.~\ref{fig:alphahscalingLam}.
In the free field theory,
a fit to the asymptotic form $\alpha_{\rm max}(\hat{r})\sim  c_1 e^{- M_\alpha \hat{r}}/\hat{r}^{3/2}$ gives
$M_\alpha=1.53(3)$ to be compared with the mass extracted from the truncated $\mathbf{K}$ matrix of $M_{\mathbf{K}} = 1.55(1)$,  two-point function of
$M_{\rm 2pt}=1.3(1)$ and inverse 2-pt function of $M_G=1.51(1)$.
The parenthetical errors express only statistical uncertainties associated with the fit.
We attribute the statistically significant differences between these masses and the input mass to finite lattice spacing and finite volume effects.
In the interacting theory, we find $M_\alpha=1.86(2)$ compared with
$M_{\rm 2pt}=1.82(1)$ and inverse 2-pt function of $M_G=1.86(1)$.
It is interesting to note that the mass of the scalar from the Hamiltonian is $M_\phi=2.22$.  This both indicates the presence of finite volume and digitization effects as well as supports the relation between the suppression of alpha angles and the correlation functions as discussed for the free field above.
Due to significant digitization errors for $n_Q=2$, the results of the calculations are sensitive to the value chosen for $\phi_{\rm max}$.
This sensitivity is expected (as seen in the tuning procedures of Ref.~\cite{Klco:2018zqz}) and  will exponentially decrease with increasing $n_Q$ once $\phi_{\rm max}$
is larger than the range of fields over which the ground state wave function has support.
In both sectors, the
classical computations (Mathematica)
converge rapidly to the lowest-lying states, and the limitation in the size
of the classical computation is imposed by the dimensionality of the underlying Hilbert space.
By performing calculations of systems with $N=4-8$ lattice sites, the ground state energy density
and single particle mass are seen to have converged to better than one percent of their infinite-volume values
(which are determined by exponential extrapolation forms).
Finite-volume systematic errors in these quantities are expected to be dictated by  $M_\phi$, and  to scale with
$\sim e^{-M_\phi N}$, along with multiplicative polynomial terms, for large lattice extents.
The $\theta$-angles and $\alpha$-angles are extracted from the resulting ground-state wave function using the methods
described previously.
After exploring the scaling with lighter masses, it is found that the results becomes less stable and more sensitive to the
choice of $\phi_{\rm max}$, though exponential suppression is retained with $m_\alpha$ values deviating from the scaling of correlation functions.
As the $\alpha$-angles are suppressed by spatial distance and finite volume effects are reduced at the expense of lattice spacing artifacts with heavy masses, it is not surprising that the heavy mass regime captures more closely the asymptotic structure of $\alpha$-angle suppression.
The $\alpha$-angles  make explicit an exponential hierarchy in control operators resulting from the
structure of the ground-state wave function.
Numerically determining the angles at large control distances requires precision in the
classical determination of the wave function at the corresponding control level.
For near term quantum computations that are relatively low precision,
only a modest number of control distances will be required, along with
comparable precision in the classical computation of the $\alpha$-angles and their truncation.

The results presented in this section demonstrate a hierarchically localized quantum circuit
preparing the ground state of an interacting field theory on a quantum register of qubits.
As with the circuit preparing a non-interacting theory, the $\alpha$-angles decrease
exponentially as a function of increasing control distance,
enabling an exponentially-precise preparation for a given control distance.
The calculations we have performed are demonstrative only.
Preparing for  simulations of scalar field theory at-scale using a quantum computer
would require more substantial classical calculations using a quantum simulator capable of assigning up to $\gsim 30$ qubits
(depending on the mass)
to accomplish the required parameter tuning,
and the determinations of the $\alpha$-angles necessary to prepare the ground state within pre-specified tolerances.

\section{Summary and Conclusions}
\noindent
Preparing the initial state of a quantum many-body system or a quantum field theory
on a quantum computer is
an essential and challenging part of quantum simulations.
To allow the subsequent processes of time evolution and measurement to be implemented straightforwardly, it is important that the initial state is prepared at high fidelity with respect to the target
wavefunction.
For scalar field theory, correlations fall exponentially with the mass of the particle as a function of space-like separation.
The mutual information, expressing the quantum and classical correlations between subsystems
(bound and distillable entanglement),
falls exponentially with twice the mass of the particle as a function of space-like separation.
Strikingly, the entanglement measure of negativity (distillable entanglement)
is found to be completely localized to the region over which the operators in the Hamiltonian are distributed.
These three features suggest that quantum circuits describing latticized scalar field theory can be
constructed with entangling operations of limited spatial extent set by the mass of the particle.
We have found one such quantum circuit design employing controlled rotation angles whose magnitudes
falls exponentially with spatial separation---consistent with the behavior of correlation functions.
This design permits truncations in the control distance and/or the $\alpha$-angles at the cost of introducing a
quantifiable degradation in the wavefunction fidelity.
While the design of the $\alpha$-circuit marks important progress in connecting quantum circuit design to correlations in the underlying physical system,  it is natural to suspect that an even more efficient quantum circuit may be possible with rotation angles that scale with twice the particle mass (or even faster) reflecting measures of the underlying entanglement.

Our results have qualitatively shown that circuit operations can be localized over a length scale that is consistent with the
lightest mass excitation in the spectrum (mass of the scalar particle), consistent with the behavior of ground-state correlation functions.
However, the demonstrated systems in both the free theory and interacting theory have substantial finite-volume effects (systematic errors),
depending upon the mass of the scalar, and significant digitization errors for two qubits per lattice site.
The latter errors, and the dependence on simulation parameters, such as the maximum value of the scalar field at each site,
are found to be substantially reduced when using three or more qubits per site.
It is encouraging that the exponential suppression of $\alpha$-angles, and thus localization of the quantum circuit, is found to be robust in the presence of these systematic errors.
While our method holds significant promise in preparing the ground state of massive, interacting field theories,
the results presented in this paper are not sufficient to inform a quantum simulation of the continuum theory with fully-quantified
uncertainties~\cite{Beane:2014oea}.
Much larger scale classical simulations,
are required with more qubits per site and more lattice sites,
in order to calculate localized $\alpha$-angles for a latticized scalar field with controlled systematic errors to be used in a simulation on a quantum computer.
We anticipate
that accomplishing these classical simulations will require (co-)developing software that can utilize heterogeneous
capability pre-exascale or exascale supercomputers.

Pseudo-scalar fields play central roles in nuclear physics, with pions providing the long-range component of the nuclear forces.
In classical numerical simulations of effective field theories describing low-energy systems of nucleons and hyperons,
dynamical pion fields remain a challenge to include at scale.
The new algorithm that we have presented here is relevant for efficiently preparing pion fields on a digital quantum computer.
Our results are anticipated to have applicability beyond one-dimensional lattice scalar field theory---aiding in circuit design for any system exhibiting a mass gap or many-body systems with significantly localized spatial correlations, including QCD and the standard model.

\appendix

\section{Continuous Lattice Scalar Field Theory with Four Spatial Sites}
\label{app:cont4sites}
\noindent
To provide  a concrete example of previous discussions, we consider a 1-dimensional system with four spatial lattice sites, $N=4$.
The allowed momenta with PBCs
are $k=0,\pm {\pi\over 2}, +\pi$, corresponding to $\hat k^2 = 0,  2 ,  2 ,  4$, and real eigenvectors
$v_i=(1,1,1,1)/2$ , $(1,0,-1,0)/\sqrt{2}$ , $(0,1,0,-1)/\sqrt{2}$ , $(1,-1,1,-1)/2$.
Transforming from momentum-space to
position-space gives a circulant matrix $\bf K$,
defined in Eq.~\eqref{eq:spuncophi},
with a first row defined by matrix elements,
\begin{align}
  \mathbf{K}_{11} &= \frac{1}{4} \left( \hat m+2 \sqrt{\hat m^2+2} + \sqrt{\hat m^2+4} \right) \nonumber \\
  \mathbf{K}_{12} = \mathbf{K}_{14} &=  \frac{1}{4} \left( \hat m- \sqrt{\hat m^2+4} \right) \nonumber \\
  \mathbf{K}_{13} &= \frac{1}{4} \left(\hat m - 2 \sqrt{\hat m^2+2} + \sqrt{\hat m^2+4} \right) \ \ \ .
  \label{eq:KcircN4}
\end{align}
The ground state wavefunction is
\begin{eqnarray}
|\psi \rangle_0  & = &
{ \hat m^{1/4} \left(\hat m^2+2\right)^{1/4}  \left(\hat m^2+4\right)^{1/8}  \over \pi}
\ \int d\phi_0 d\phi_1 d\phi_2 d\phi_3 \
| \phi_0 \phi_1 \phi_2 \phi_3 \rangle\
\nonumber\\
&&
e^{-{1\over 2} \left( {\bf K}_{11} ( \phi_0^2+ \phi_1^2+ \phi_2^2+ \phi_3^2)
+ 2 {\bf K}_{12} \left( \phi_0 \phi_1+\phi_1 \phi_2+\phi_2 \phi_3+\phi_3 \phi_0 \right)
+ 2 {\bf K}_{13} \left( \phi_0 \phi_2+\phi_1 \phi_3 \right)
\right)
}
\ \ \ .
\label{eq:psi4}
\end{eqnarray}
Ground-state expectation values of bi-linears are
\begin{eqnarray}
_0\langle \psi  | \ \phi(j) \phi(j)\  | \psi \rangle_0
& = &
{1\over 8}\left( {1\over \hat m} + {2\over\sqrt{\hat m^2+2}} + {1\over\sqrt{\hat m^2+4}} \right)
\nonumber\\
_0\langle \psi  | \ \phi(j) \phi(j+1)\  | \psi \rangle_0
& = &
{1\over 8}\left( {1\over \hat m} - {1\over\sqrt{\hat m^2+4}} \right)
\nonumber\\
_0\langle \psi  | \ \phi(j) \phi(j+2)\  | \psi \rangle_0
& = &
{1\over 8}\left( {1\over \hat m} - {2\over\sqrt{\hat m^2+2}} + {1\over\sqrt{\hat m^2+4}} \right)
\ \ \ .
\label{eq:phiphiN4}
\end{eqnarray}
%

\section{Mutual Information}
\label{app:MI}

The density matrix given in Eq.~(\ref{eq:reduced12K}) can be readily reduced into a
manageable form as all integrations
are Gaussian~\cite{Srednicki:1993im}.
The ${\rm\bf K}$ matrix in Eq.~\eqref{eq:spuncophi} is  partitioned in a way to isolate
the $\bm\phi$ target sites from the traced degrees of freedom $\bar{\bm\phi}$,
leading to a reduced density matrix of the form
\begin{eqnarray}
\hat{\bm \rho}  ({\bm\phi},{\bm\phi}')
& \propto &
\exp \left[ -\frac{1}{2}  \left( \bm\phi^T {\bm \gamma} \bm\phi + \bm\phi'^T  {\bm \gamma} \bm\phi' \right)
+ \bm \phi^T   {\bm\beta} \bm\phi' \right]
\ \ \ ,
\nonumber\\
  {\bm \gamma} & = & C -  {\bm\beta}
\ \ ,\ \
  {\bm \beta }= \frac{1}{2} B^T A^{-1} B
\ \ ,\ \
 \mathbf{K} \ = \   \left(
\begin{array}{c|c}
 A & B \\
 \hline
 B^T & C \\
\end{array}
\right)
\label{eq:rho1gbAPP}
  \ \ \ .
\end{eqnarray}
No assumptions have  been made about the boundary conditions imposed on the lattice fields
or structure of the correlations.
The diagonal  terms can be decoupled by a change of variables,
\begin{equation}
   {\bm\gamma} = V^T \gamma_D V \ \ , \ \
  \tilde{\bm \phi} = \gamma_D^{\frac{1}{2}} V \bm\phi \ \ , \ \
  \tilde{\bm \phi}^T = \bm\phi V^T \gamma_D^{\frac{1}{2}} \ \ , \\
  \label{eq:diagonalizationTransform}
\end{equation}
for both
${\bm\phi}$ and ${\bm\phi}'$ resulting in
\begin{equation}
  \hat{\bm \rho} ({\bm\phi},{\bm\phi}')
  \propto
  \exp \left[-\frac{1}{2} \left( \tilde{\bm\phi}^T\tilde{\bm\phi} + \tilde{\bm\phi}'^T \tilde{\bm\phi}' \right)
  + \tilde{\bm\phi}^T  {\bm\beta}' \tilde{\bm\phi}' \right]
  \ \ ,
  \qquad
   {\bm\beta}' = \gamma_D^{-\frac{1}{2}} V  {\bm\beta} V^T \gamma_D^{-\frac{1}{2}}
  \ \ \ .
  \label{eq:betaprimeRHO}
\end{equation}
The eigenvalues of $ \hat{\bm \rho} ({\bm\phi},{\bm\phi}')$ are  characterized by the eigenvalues
of $ {\bm\beta}'$, denoted by $\beta'_i$.
The number of $\beta'_i$ corresponds to the dimensionality of the number of lattice sites in the
subsystem, and the
the eigenvalues of $\hat{\bm\rho} ({\bm\phi},{\bm\phi}')$ are  multiplicative combinations of the
towers of eigenvalues for the individual oscillators, as given in Eq.~(\ref{eq:RHO1eigenvals}).
In the case of the reduced space corresponding to two spatial sites, the $ {\bm\beta}'$ matrix is $2\times 2$ dimensional,
with eigenvalues
\begin{eqnarray}
\beta'_1 & = & {  \beta_{11}-\beta_{12}  \over \gamma_{11}-\gamma_{12} }
\ \ ,\ \
\beta'_2 \ =\  {  \beta_{11}+\beta_{12}  \over \gamma_{11}+\gamma_{12} }
  \ \ \ ,
  \label{eq:betapAPPB}
\end{eqnarray}

Returning to Eq.~(\ref{eq:rho1gbAPP}), it is illuminating to reveal the lattice site dependence explicitly.
Restricting the reduced space to only two spatial sites, $0$ and $j$,
the argument of the exponential in $ \hat{\bm \rho} ({\bm\phi},{\bm\phi}')$ is
\begin{eqnarray}
{\cal A} & = &
 \bm\phi^T  {\bm\gamma} \bm\phi + \bm\phi'^T  {\bm\gamma} \bm\phi'   -2 \bm \phi^T  {\bm\beta} \bm\phi'
 \nonumber\\
 & = & \gamma_{11} \left( \phi_0^2 + \phi_0'^2+ \phi_j^2 + \phi_j'^2 \right)
 \ -\ 2 \beta_{11}  \left(  \phi_0\phi_0' +  \phi_j\phi_j' \right)
 \nonumber\\
 & &
 \ +\ 2\gamma_{12}  \left(  \phi_0\phi_j +  \phi_0'\phi_j' \right)
 \ -\ 2\beta_{12}  \left(  \phi_0\phi_j' +  \phi_0'\phi_j \right)
 \ \ ,
  \label{eq:GammaAPP}
\end{eqnarray}
where $\gamma_{11}, \beta_{11}$ are numbers of order unity with slight separation dependence,
but $\gamma_{12}, \beta_{12} \sim e^{-m j} /j^p$ can be shown to fall faster than exponentially
with separation, where $p$ is a positive integer.
In the limit $j\rightarrow \infty$ with $\gamma_{12}, \beta_{12} =0$, $ \hat{\bm \rho} ({\bm\phi},{\bm\phi}')$
can be written as a tensor product of
$ \hat{\bm \rho} ({\bm\phi},{\bm\phi}') = \hat{\boldsymbol{\rho}}_0 (\phi_0,\phi_0')\otimes  \hat{\boldsymbol{\rho}}_0 (\phi_j, \phi_j')$,
where
\begin{eqnarray}
\hat \rho_0 ( \phi,\phi') & \propto &
 \exp \left[\  -{1\over 2} {\bm\Phi}^T  {\bm\Sigma}_\infty {\bm\Phi} \ \right]
 \nonumber\\
 {\bm\Phi}^T & = & \left( \phi, \phi'\right)
 \ \ ,\ \
 {\bm \Sigma}_\infty \ =\
 \left(
 \begin{array}{cc}
 \gamma_{11} & -\beta_{11} \\
- \beta_{11} &  \gamma_{11}
 \end{array}
 \right)
 \ \ ,
  \label{eq:singleAPP}
\end{eqnarray}
as the $\beta_i'$ in Eq.~(\ref{eq:betapAPPB}) become degenerate.
Making explicit the process discussed in Ref.~\cite{Srednicki:1993im},
 an eigenstate of an harmonic oscillator is sought, along with its angular frequency,
\begin{eqnarray}
&& \psi_n (\phi) \ \propto \
H_n(\sqrt{\alpha} \phi) e^{-{1\over 2}\alpha \phi^2}
\ \ ,\ \
\int\ d\phi'\ \hat \rho_0 ( \phi,\phi') \ \psi_n (\phi') \ =\  \lambda_{n} \psi_n (\phi)
 \ \ ,
  \label{eq:solvingpsiAPP}
\end{eqnarray}
which requires $\alpha=\sqrt{\gamma_{11}^2-\beta_{11}^2}$, with eigenvalues
\begin{eqnarray}
\lambda_{n}  & = & {\gamma_{11}-\beta_{11}+\alpha \over \gamma_{11}+\alpha}
\left({\beta_{11}\over \gamma_{11}+\alpha}\right)^n
 \ \ ,
  \label{eq:evalsinftyAPP}
\end{eqnarray}
which automatically satisfies ${\rm Tr}\left[\ \hat{\boldsymbol{\rho}}_0 \ \right]=1$.
The eigenstates take  the form
\begin{eqnarray}
\psi_{n_0, n_j} (\phi_0, \phi_j)
& = &  \psi_{n_0} (\phi_0)   \psi_{n_j} (\phi_j)
 \ \ ,
  \label{eq:psi2APP}
\end{eqnarray}
which has an eigenvalue $ \lambda_{n_0}  \lambda_{n_j}  $ when $\gamma_{12}, \beta_{12} =0$,
and eigenvalues that can be computed in perturbation theory for
$\gamma_{12}, \beta_{12} \sim e^{-m j}/j^p$ as $j\rightarrow \infty$.
We examine first-order perturbation theory in $\gamma_{12}, \beta_{12}$.
Expanding the exponential function in Eq.~(\ref{eq:GammaAPP}), the fields enter linearly,
and hence do not contribute to diagonal matrix elements, leaving off-diagonal contributions only.
When connecting spaces with different unperturbed matrix elements, they contribute only
quadratic changes to the eigenvalues.  In the case of off-diagonal matrix elements in
degenerate spaces, such as the configurations $(n_0,n_j)=(1,2)$ and $(2,1)$,
diagonalization provides eigenvalues that are linearly shift by $\gamma_{12}, \beta_{12}$,
both positively and negatively.
When computing the MI using Eq.~(\ref{eq:summedS}), the sum over these eigenvalues
provides a contribution that is also quadratic in
$\gamma_{12}, \beta_{12}$.
Therefore, at large separations, the MI scales as $\sim e^{-2 m j}/j^{p'}$ as $j\rightarrow \infty$,
falling exponentially faster than the two-point correlations.

\section{Negativity}
\label{app:nega}

The negativity provides information about the entanglement of a quantum field theory that complements
that from correlation functions and MI.  In this Appendix we provide a simple explicit
example, and also develop the framework for determining $\mathcal{N}(\hat{\bm \rho})$ for a lattice scalar field theory.

\subsection{A Lattice with Two Sites}
\label{app:negaTWO}

In calculating the negativity between two sites in a two-site lattice, $\mathcal{N}(\hat{\bm \rho})$,
the Hilbert space of the first site is partially transposed without any  Hilbert space being removed through tracing.
The partial transpose of the two-site ground state density matrix over the degrees of
freedom in the first site, $\hat {\bm \rho}^\Gamma$,
is related to the density matrix of the system $\hat {\bm \rho}$
\begin{eqnarray}
 \hat {\bm \rho} & = &
 \int \dif \phi_1 \dif \phi_1' \dif \phi_2 \dif \phi_2' \ \
 \langle \phi_1', \phi_2' | \psi\rangle_0
  {}_0\langle \psi | \phi_1, \phi_2 \rangle\
 \ |\phi_1', \phi_2'\rangle  \langle \phi_1, \phi_2|
\nonumber\\
 \hat {\bm \rho}^\Gamma & = &
 \int \dif \phi_1 \dif \phi_1' \dif \phi_2 \dif \phi_2' \ \
 \langle \phi_1, \phi_2' | \psi\rangle_0
  {}_0\langle \psi | \phi_1', \phi_2 \rangle\
 \ |\phi_1', \phi_2'\rangle  \langle \phi_1, \phi_2|
 \ \ \ ,
\end{eqnarray}
where the (real) wavefunction has been defined in Eq.~(\ref{eq:spuncophi}).
It is straightforward to show that
the eigenvectors of $\hat {\bm \rho}^\Gamma$ are symmetric and anti-symmetric combinations of harmonic oscillator wavefunctions
\begin{equation}
  \psi^\pm_{n_1, n_2}(\phi_1, \phi_2)
  \propto
  \frac{H_{n_1}( \sqrt{\alpha} \phi_1 ) H_{n_2} ( \sqrt{\alpha} \phi_2 ) \pm
  H_{n_2}( \sqrt{\alpha} \phi_1 ) H_{n_1}( \sqrt{\alpha} \phi_2 )}{\sqrt{2}}
  e^{- \frac{\alpha}{2} \left( \phi_1^2 + \phi_2^2\right)}
  \ \ \ ,
  \label{eq:partialtransposerhoN2}
\end{equation}
for $n_1\ne n_2$ and
\begin{equation}
  \psi^\pm_{n, n}(\phi_1, \phi_2)
  \propto
 H_{n}( \sqrt{\alpha} \phi_1 ) H_{n} ( \sqrt{\alpha} \phi_2 )
  e^{- \frac{\alpha}{2} \left( \phi_1^2 + \phi_2^2\right)}
  \ \ \ ,
  \label{eq:partialtransposerhoN2eq}
\end{equation}
for $n_1=n_2=n$.
The eigenvalues associated with $ \psi^\pm_{n, n}(\phi_1, \phi_2) $ are positive,
and hence do not contribute to
$\mathcal{N}( \hat {\bm \rho})$. In contrast,
the eigenvalues associated with $\psi^\pm_{n_1, n_2}(\phi_1, \phi_2)$
in Eq.~(\ref{eq:partialtransposerhoN2}) appear in pairs of equal magnitude and opposite
sign for the two  states of definite symmetry,
\begin{equation}
  \hat {\bm \rho}^\Gamma \ \psi^\pm_{n_1,n_2} \left( \phi_1, \phi_2\right)
  =
  \pm \left( \frac{   \sqrt{\det {\bf K}}  }{\pi^{N/2}} \right)
  \frac{2 \pi( -{\bf K}_{12} )^{n_1+n_2} }{({\bf K}_{11}+\alpha)^{n_1+n_2+1} }
  \ \psi^{\pm}_{n_1, n_2} \left( \phi_1, \phi_2\right)
  \ \ , \qquad \alpha =  \sqrt{\det {\bf K}}
  \ \ \ .
  \label{eq:ptevalN2}
\end{equation}
After summing the negative eigenvalues, the negativity is found to be
\begin{equation}
  \mathcal{N} (\hat{\bm \rho}) = \sum_{\lambda_i<0} |\lambda_i|
  = -\frac{1}{2} + \sqrt{\det{\bf K}} \sum_{n_1,n_2 = 0}^\infty
  \frac{|{\bf K}_{12}^{n_1+n_2}|}{( {\bf K}_{11}+\alpha)^{n_1+n_2+1}}
  = \frac{|{\bf K}_{12}|}{{\bf K}_{11}-|{\bf K}_{12}|+\alpha}
  \ \ \ .
  \label{eq:negativityN2APP}
\end{equation}
As expected, $ \mathcal{N} (\hat{\bm \rho}) $
vanishes when ${\bf K}_{12}=0$ and the two sites decouple to produce independent fields.
The sign of ${\bf K}_{12}$ can be seen in Eq.~(\ref{eq:ptevalN2}) to be a contributing factor in the assignment
of the positive or negative eigenvalues to the symmetric or anti-symmetric wavefunctions.
The negativity is naturally independent of this assignment.

\subsection{Lattices with more than Two Sites}

Considering the negativity between two lattice sites, $0$ and $j$,
the computation of $ \mathcal{N} (\hat{\bm \rho}_{0j}) $ closely follows that of the MI, after integrating over the other sites,
the argument of the exponential in Eq.~(\ref{eq:GammaAPP})  is transformed to
\begin{eqnarray}
{\cal A}_{0j} \rightarrow {\cal A}_{0j}^\Gamma & = &
  \gamma_{11} \left( \phi_0^2 + \phi_0'^2+ \phi_j^2 + \phi_j'^2 \right)
 \ -\ 2 \beta_{11}  \left(  \phi_0\phi_0' +  \phi_j\phi_j' \right)
 \nonumber\\
 & &
 \ +\ 2\gamma_{12}  \left(  \phi_0'\phi_j +  \phi_0\phi_j' \right)
 \ -\ 2\beta_{12}  \left(  \phi_0'\phi_j' +  \phi_0\phi_j \right)
 \ \ ,
  \label{eq:AAPP}
\end{eqnarray}
which can be included in subsequent calculations by a definition of the $\gamma\rightarrow\gamma^\Gamma$ and
$\beta\rightarrow\beta^\Gamma$ matrices in Eq.~(\ref{eq:GammaAPP}), by $\gamma_{12}\leftrightarrow -\beta_{12}$.
The system is once again reduced to two uncoupled oscillators.
The eigenvalues of the $ {\bm\beta}'\rightarrow  {\bm\beta}^{\Gamma '}$ matrix are
\begin{eqnarray}
 \beta^{'\Gamma}_1 & = & {  \beta_{11}-\gamma_{12}  \over \gamma_{11}-\beta_{12} }
\ \ ,\ \
\beta^{'\Gamma}_2 \ =\  {  \beta_{11}+\gamma_{12}  \over \gamma_{11}+\beta_{12} }
  \ \ \ ,
  \label{eq:betapAPP}
\end{eqnarray}
leading to eigenvalues of
$\hat{\bm \rho}_{0j}^\Gamma$
via Eq.~(\ref{eq:RHO1eigenvals}).
If the negativity is non-zero, one of these oscillators is without negativity ($\xi_1 > 0$)
while the other has eigenvalues of alternating sign $(\xi_2 < 0)$.
In such a case, the negativity calculation may be simplified to
\begin{equation}
  \mathcal{N}(\hat {\bm \rho}_{0j}) = -\sum_{n=0}^\infty \sum_{m=0}^\infty \lambda_{1,n} \lambda_{2,2m+1}
  = - \frac{ \xi_2}{1+\xi_2} \ \ \ ,
\end{equation}
equal to the negativity of the second decoupled oscillator.

\section{Digitized Lattice Scalar Field Theory with Two Spatial Sites}
\label{app:Dig22sites}
\begin{figure}[!ht]
	\centering
	\includegraphics[width=0.75\columnwidth]{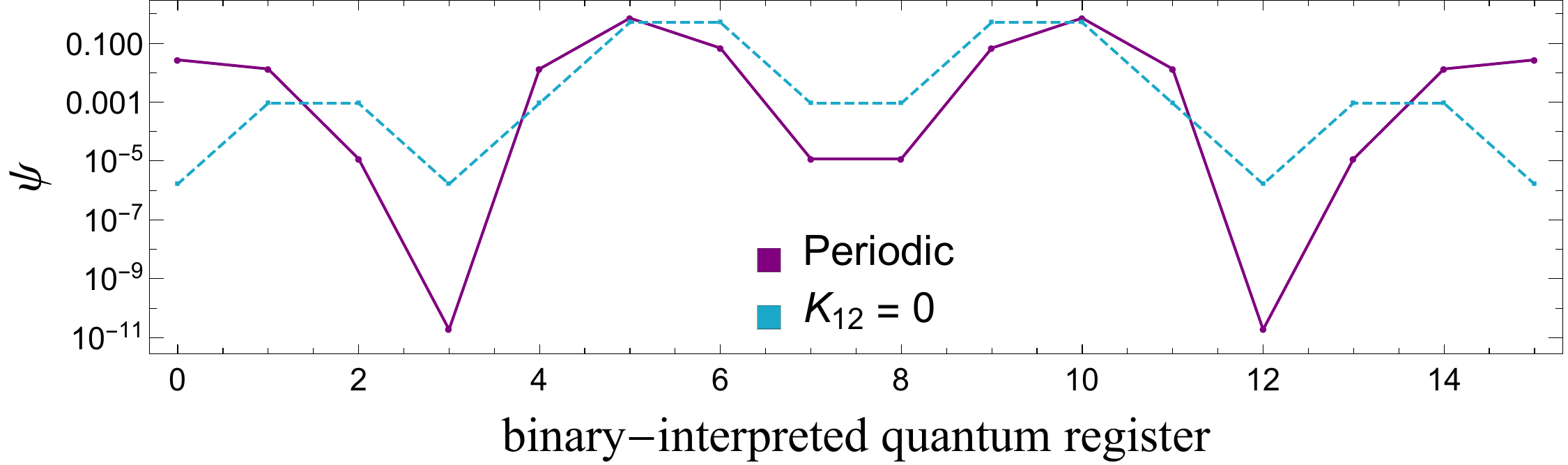}
	\caption{
	The ground state wavefunction for a digitized, free lattice scalar field theory with
	$\hat{m}=0.3 $ calculated on $N=2$ spatial sites (purple points) from Eq.~(\ref{eq:spuncophiOPact22}).
	The blue points (dashed line) correspond to setting ${\bf K}_{12}=0$, while the purple points (solid line) correspond to the full entangled wave function.
		}
		\label{fig:psi22}
\end{figure}

\noindent
It is helpful to consider the simplest non-trivial digitized scalar field theory, one with two spatial sites, $N=2$, with
two qubits per site, $n_Q=2$.
The scalar field at each site is digitized into a four-dimensional Hilbert space, with states defined by
$\{\   |-\phi_{\rm max}\rangle ,  |-{1\over 3}\phi_{\rm max}\rangle  ,  |+{1\over 3}\phi_{\rm max}\rangle  ,  |+\phi_{\rm max}\rangle \ \}$.
The basis for the field theory is the tensor product of this basis at the two sites, an element of which we denote by, for example,
$ |-{1\over 3}\phi_{\rm max}\rangle_1\otimes  | + \phi_{\rm max}\rangle_2 \rightarrow | -{1\over 3}, +1 \rangle$.
The operator defined in Eq.~(\ref{eq:spuncophiOP}) becomes,
\begin{eqnarray}
\hat \Gamma & = & e^{- {1\over 2} \left( {\bf K}_{11} ( \hat\phi_1^2  +  \hat\phi_2^2 ) + 2 {\bf K}_{12} \hat\phi_1\hat\phi_2 \right) }
\ \ \ ,
\label{eq:spuncophiOP22}
\end{eqnarray}
where use has been made of the circulant structure and rotational symmetry of ${\bf K}$.
The action of this operator gives an entangled wave function of the form
\begin{eqnarray}
|\psi_d\rangle &  = & A\
\left[\
 e^{- \phi_{\rm max}^2 \left(  {\bf K}_{11} + {\bf K}_{12} \right)}
 \left[ \ |+1,+1\rangle + |-1,-1\rangle\ \right] \right. \nonumber \\
&&
 \ +\
 e^{- \phi_{\rm max}^2 \left( {\bf K}_{11} - {\bf K}_{12} \right)}\
  \left[ \ |+1,-1\rangle + |-1,+1\rangle\ \right]
 \nonumber\\
&&
 \ +\
 e^{- \phi_{\rm max}^2 \left( {5\over 9} {\bf K}_{11} - {1\over 3}{\bf K}_{12} \right)}\
  \left[\
 |+1,-{1\over 3}\rangle + |-1,+{1\over 3}\rangle +  |-{1\over 3}, +1\rangle + |+{1\over 3}, -1\rangle
\right]
 \nonumber\\
&&
 \ +\
 e^{- \phi_{\rm max}^2 \left( {5\over 9} {\bf K}_{11} + {1\over 3}{\bf K}_{12} \right)}\
  \left[\
 |+1,+{1\over 3}\rangle + |-1,-{1\over 3}\rangle +  |-{1\over 3}, -1\rangle + |+{1\over 3}, +1\rangle
\right]
\nonumber\\
&&
 \ +\
 e^{- \phi_{\rm max}^2 {1\over 9} \left( {\bf K}_{11} + {\bf K}_{12} \right)}
 \left[ \ |+{1\over 3},+{1\over 3}\rangle + |-{1\over 3},-{1\over 3}\rangle\ \right] \nonumber \\
 && \left.\ +\
 e^{- \phi_{\rm max}^2 {1\over 9} \left( {\bf K}_{11} - {\bf K}_{12} \right)}
 \left[ \ |+{1\over 3},-{1\over 3}\rangle + |-{1\over 3},+{1\over 3}\rangle\ \right]
\ \right]
\ \ \ ,
\label{eq:spuncophiOPact22}
\end{eqnarray}
where the normalization factor is
\begin{eqnarray}
A^{-2} & = &
2  e^{- 2 \phi_{\rm max}^2 \left( {\bf K}_{11} + {\bf K}_{12} \right)}
\ + \
2  e^{- 2 \phi_{\rm max}^2 \left( {\bf K}_{11} - {\bf K}_{12} \right)}
\ +\
4 e^{- 2\phi_{\rm max}^2 \left( {5\over 9} {\bf K}_{11} + {1\over 3}{\bf K}_{12} \right)}
 \nonumber\\
& + &
4 e^{- 2\phi_{\rm max}^2 \left( {5\over 9} {\bf K}_{11} - {1\over 3}{\bf K}_{12} \right)}
\ +\
2 e^{- 2\phi_{\rm max}^2 {1\over 9} \left( {\bf K}_{11} + {\bf K}_{12} \right)}
\ +\
2 e^{- 2\phi_{\rm max}^2 {1\over 9} \left( {\bf K}_{11} - {\bf K}_{12} \right)}
\ \ \ .
\label{eq:spuncophiOPact22Anorm}
\end{eqnarray}
The symmetry under ${\bm\phi}\rightarrow -{\bm\phi}$ is explicit, as is the symmetry $\phi_1\leftrightarrow\phi_2$ of the two-site system.
The elements of the first row of the ${\bf K}$ matrix are,
\begin{eqnarray}
{\bf K}_1 &  = &
{1\over 2} \left( \hat{m} + \sqrt{\hat{m}^2+4} , \hat{m}- \sqrt{\hat{m}^2+4}
\right)
\ \ \ .
\label{eq:KcircN2}
\end{eqnarray}
As a numerical example, Fig.~\ref{fig:psi22} shows this wavefunction evaluated for $\hat{m}=0.3$ and $\phi_{\rm max}=3.5$,
and also for the case with ${\bf K}_{12}=0$.
The extension to systems with more sites follows straightforwardly from this example.
The two-site system does not capture the richness of the correlation and
entanglement structure of larger systems, particularly the fall off
with increasing separation between sites.

\section{$\theta$- and $\alpha$-Distributions at Fixed Control Distance $h$}
\label{app:anglesfixedcontrol}
\begin{figure}
  \includegraphics[width = 0.46\textwidth]{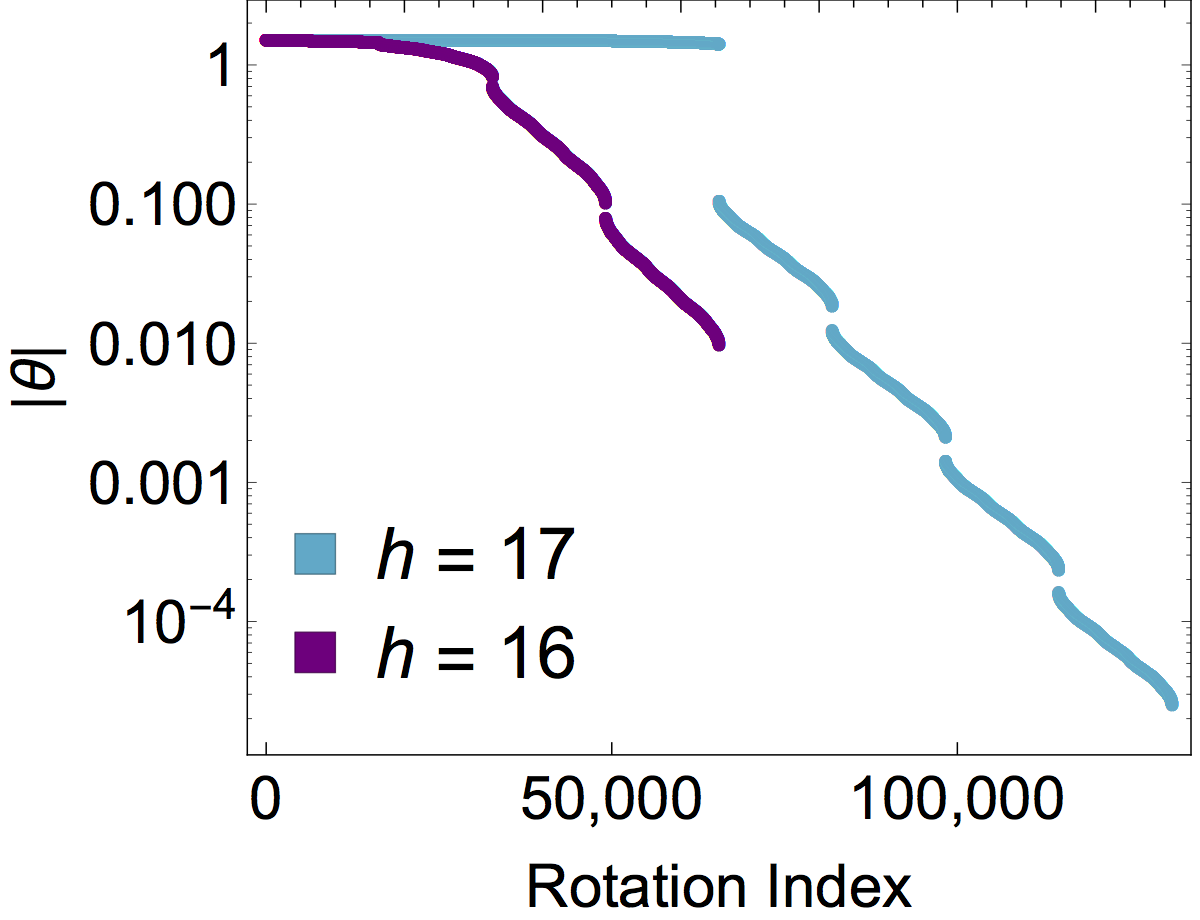}
  \includegraphics[width = 0.45\textwidth]{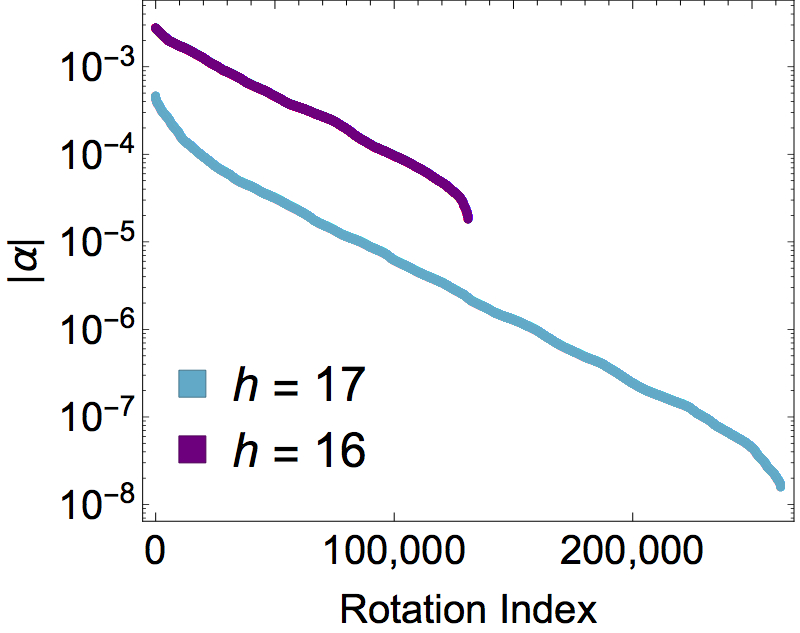}
  \caption{Ordered magnitudes of the $\theta$-angles (left) and $\alpha$-angles (right) associated with control distances $h=16, 17$.   The behaviors of these angles at these control distances are representative of those at other distances.}
  \label{fig:angleRotIndex}
\end{figure}

In Fig.~\ref{fig:alphahscaling}, it is shown that the physically-motivated $\alpha$ angles have been endowed with a hierarchy aligned with that of the $\mathbf{K}$ matrix defined in Eq.~(\ref{eq:spuncophiOP}), decaying exponential with physical separation.
This hierarchy is most notably expressed in the progression of the maximum magnitude of the angles as a function of $h$, exponentially decaying with the scalar mass.
However, an additional advantageous feature is embedded in the $\alpha$-angles at fixed $h$. This is demonstrated in
Fig.~\ref{fig:angleRotIndex}, where the angles associated with rotation operators of $h = 16,17$ that are used to initialize a ten-site scalar field with two qubits per site are ordered by magnitude.  Though it is depicted as roughly a single point in Fig.~\ref{fig:alphahscaling},  Fig.~\ref{fig:angleRotIndex} shows that approximately half of the $\theta$-angles at each level, $\ell= h$, reside around $\theta \sim \frac{\pi}{2}$.
As this structure is representative of that at every value of $h$,
the number of operators characterized by $\mathcal{O}(1)$ rotation angles scales exponentially with the number of controls on the operator, $2^{h-1}$.
Highly non-local operators dominate the number of entangling operations required for state preparation.
In contrast, the $\alpha$-angles fall continuously at fixed $h$, in addition to the $h$-dependent suppression shown in the main text.

\section{Site-wise $\alpha$-Angles}
\label{app:AlphaExample}

In the main text, a recursive algorithm for generating  site-wise
$\alpha$-angles from the corresponding $\theta$-angles has been presented in  Eq.~(\ref{eq:alphasRec}).
In this Appendix, an explicit example is provided of such a site-wise decomposition, along with numerical values of the resulting $\alpha$-angles.

The quantum circuit transformation that will be utilized in this appendix is
\begin{multline}
  \begin{gathered}
    \scalebox{0.8}{
   \Qcircuit @C=0.2em @R=0.6em {
   &  \gate{R(\theta_0)} & \ctrlco{1} & \ctrlco{1} & \ctrlco{1} &  \qw &   \ctrlco{1} & \ctrlco{1} & \qw \\
   & \qw & \gate{R(\vec{\theta}_1)} & \ctrlco{1} & \ctrlco{1} & \qw &\ctrlco{1} & \ctrlco{1} & \qw\\
   & \qw &  \qw & \gate{R(\vec{\theta}_2)} & \ctrlco{1} & \qw & \ctrlco{1} & \ctrlco{1} & \qw\\
   & \qw &  \qw & \qw & \gate{R(\vec{\theta}_3)} & \qw & \ctrlco{1} & \ctrlco{1} & \qw \\
   & \qw & \qw & \qw & \qw & \qw & \gate{R(\vec{\theta}_4)} & \ctrlco{1} & \qw \\
   & \qw & \qw & \qw & \qw & \qw & \qw & \gate{R(\vec{\theta}_5)} & \qw \gategroup{3}{1}{4}{9}{2.4em}{--}
   }
   }
   \end{gathered} \qquad
   =
   \\
   \vspace{0.2cm}
   \begin{gathered}
     \scalebox{0.8}{
     \Qcircuit @C=0.2em @R=0.6em {
     & \gate{R(\alpha_{00})} & \ctrlco{1} & \ctrlco{1} & \qw &  \ctrlco{1} & \qw & \ctrlco{1} & \qw & \qw & \ctrlco{1} & \qw  \\
     & \qw & \gate{R(\vec{\alpha}_{11})} & \ctrlco{1} & \qw & \ctrlco{1} & \qw & \ctrlco{1} & \qw & \qw & \ctrlco{1} & \qw \\
     & \qw & \gate{R(\alpha_{20})} & \gate{R(\vec{\alpha}_{22})} & \ctrlco{1} & \ctrlco{1} &  \ctrlco{1} & \ctrlco{1} & \qw & \ctrlco{1} & \ctrlco{1} & \qw \\
     & \qw & \qw & \qw & \gate{R(\vec{\alpha}_{31})} & \gate{R(\vec{\alpha}_{33})} &  \ctrlco{1} &  \ctrlco{1} & \qw & \ctrlco{1} & \ctrlco{1} & \qw \\
     & \qw & \qw & \qw & \qw & \gate{R(\alpha_{40})} & \gate{R(\vec{\alpha}_{42})} & \gate{R(\vec{\alpha}_{44})} & \ctrlco{1} & \ctrlco{1} & \ctrlco{1} & \qw \\
     & \qw & \qw & \qw & \qw & \qw & \qw & \qw & \gate{R(\vec{\alpha}_{51})} & \gate{R(\vec{\alpha}_{53})} & \gate{R(\vec{\alpha}_{55})} & \qw  \gategroup{3}{1}{4}{12}{2.2em}{--}
     }
     }
   \end{gathered} \ \ \ .
   \label{eq:explicitcircuit}
\end{multline}
This particular site-wise decomposition is designed for Hilbert spaces defined by two qubits, $n_Q = 2$, at each of three spatial lattice sites, $N = 3$.
The dashed rectangle encloses the second lattice site for clarity.
Eq.~\eqref{eq:explicitcircuit}
shows the site-wise arrangement of qubits and the decomposition of the  $\theta$-angle control operators.
As discussed in the text surrounding Eq.~\eqref{eq:alphasRec}, there exists a choice in defining the site-wise $\alpha$-angles.  While the operator of length $h = \ell$ must be retained at each level, any set of additional localized operators can be introduced at each level.  Here, a choice is made such that a single operator is included at each physical distance, $r$, and at each $r$ the operator with the largest $h$ is introduced.  This adds dependence on (entire) neighboring sites, inspiring the description of this protocol as transforming to site-wise $\alpha$-angles.

Consider the controlled operations initializing the $6^{\rm th}$ qubit ($\ell = 5$), accomplished
with a five-qubit-controlled operator ($h=5$) including qubits from all three sites, defined by the
$\theta_{5,k}$ angles where $k=0,1, .. ,2^5-1$.
This operator is decomposed into three different controlled operators of depth $h=1,3,5$
(i.e. there is no single qubit rotation,
which would correspond to $h=0$, or split-site operators which would have $h=2,4$).
This is a particular choice of decomposition that is physically motivated by the underlying quantum field theory.
The numerical computations to be performed are the evaluations of $\alpha_{51,k}$, $\alpha_{53,k}$
and $\alpha_{55,k}$,
\begin{eqnarray}
\alpha_{51,k} & = & \sum_{m=0}^{15}\ {\theta_{5,k+2 m} \over 16}
\ \ \ {\rm with}\ \ \ k=0,1
\nonumber\\
\alpha_{53,k} & = & \sum_{m=0}^{3}\ {\theta_{5,k+8 m} \over 4}
\ -\
\alpha_{51,k\bmod 2}
\ \ \ {\rm with}\ \ \ k=0,1, ... , 7
\nonumber\\
\alpha_{55,k} & = &  \theta_{5,k}
\ -\
\alpha_{51,k\bmod 2}
\ -\
\alpha_{53,k\bmod 8}
\ \ \ {\rm with}\ \ \ k=0,1, ... , 31
\ \ \ .
\label{eq:alphal5odd}
\end{eqnarray}
Similarly, the preparation of the $5^{\rm th}$ qubit (which does have an uncontrolled single qubit rotation in this chosen $\alpha$-decomposition)
is accomplished
with an operator with four controls ($h=4$) including qubits from all three sites, defined by the
$\theta_{4,k}$ angles where $k=0,1, .. ,15$.
This operator can be decomposed into
one single qubit rotation and  two  controlled operators, of control distances $h=2,4$.
The numerical computations to be performed are evaluations of $\alpha_{40,k}$, $\alpha_{42,k}$
and $\alpha_{44,k}$,
\begin{eqnarray}
\alpha_{40} & = & \sum_{m=0}^{15}\ {\theta_{4,m} \over 16}\ =\ {\pi\over 4}
\nonumber\\
\alpha_{42,k} & = & \sum_{m=0}^{3}\ {\theta_{4,k+4 m} \over 4}
\ -\
\alpha_{40}
\ \ \ {\rm with}\ \ \ k=0, ... , 3
\nonumber\\
\alpha_{44,k}
& = &
\theta_{4,k}
-
\alpha_{42,k\bmod 4}
-
\alpha_{40}
\ = \
 \theta_{4,k}
-
\sum_{m=0}^{3}\ {\theta_{4,4 m + k\bmod 4} \over 4}
\ \ \ {\rm with}\ \ \ k=0, ... , 15
\ .
\label{eq:alphal4even}
\end{eqnarray}
The angles have been grouped to make clear the suppression with respect to spatial separation. Explicitly for this example,
the angles are grouped in the following way:
$r=0$: (l=1,3,5 with h=1) and (l=0,2,4 with h=0),
$r=1$: (l=3,5 with h=3) and (l=2,4 with h=2),
$r=2$: (l=5 with h=5) and (l=4 with h=4).
Numerical results for the $\theta$- and $\alpha$-angles from a free field with open boundary conditions and ground state wavefunction defined by
\begin{equation}
  \mathbf{K} = \left(
\begin{array}{ccc}
 1.396 & -0.371 & -0.0493 \\
 -0.371 & 1.347 & -0.371 \\
 -0.0493 & -0.371 & 1.396 \\
\end{array}
\right) \ \ \ ,
\end{equation}
are shown in
Table~\ref{tab:thetaalphaeg} as a function of the spatial separation, $r$.  The parameters used to define the latticized field in this example are: $N=3$, $n_Q=2$, $m=0.3$, and $\phi_{\rm max}=3.5$.

\begin{table}
  \begin{tabular}{c|c|c|c|c}
\hline
\hline
  $\ell$ & $\vec{\theta}_\ell$ & $\alpha(\hat{r} = 0)$ & $\alpha(\hat{r} = 1)$ & $\alpha(\hat{r} = 2)$ \\
\hline
\hline
  0 & 0.7854 & 0.7854 & &  \\
  \hline
  1 & 1.569, 0.001463 & 1.569, 0.001463 &  &   \\
  \hline
  2 & \begin{tabular}{cc} 0.03564 & 0.3174 \\1.253 & 1.535 \end{tabular} & 0.7854 & \begin{tabular}{cc} -0.7498 & -0.468 \\ 0.468 & 0.7498 \end{tabular} &  \\
  \hline
  3 & \begin{tabular}{cc} 1.535 & 0.00007622 \\1.566 & 0.0006039 \\1.570 & 0.004701 \\1.571 & 0.03614 \end{tabular} & 1.56, 0.01038 & \begin{tabular}{cc} -0.02576 & -0.01030 \\0.005679 & -0.009776 \\0.009776 & -0.005679 \\0.01030 & 0.02576 \end{tabular} &  \\
  \hline
  4 & \begin{tabular}{cc}
     0.03223 & 0.2387 \\1.072 & 1.499 \\0.04216 & 0.3082 \\1.176 & 1.516 \\0.05512 & 0.3945 \\1.263 & 1.529 \\0.07206 & 0.4988 \\1.332 & 1.539 \\
   \end{tabular}& 0.7854 &\begin{tabular}{cc} -0.735 & -0.4254 \\ 0.4254 &  0.735 \end{tabular}
   & \begin{tabular}{cc}
   -0.01816 & -0.1214 \\-0.1387 & -0.02167 \\-0.008237 & -0.05188 \\-0.03451 & -0.004731 \\0.004731 & 0.03451 \\0.05188 & 0.008237 \\0.02167 & 0.1387 \\0.1214 & 0.01816 \\
   \end{tabular} \\
  \hline
  5 & \begin{tabular}{cc} 1.555 & 0.00001610 \\1.569 & 0.0001215 \\1.571 & 0.0009167 \\1.571 & 0.006916 \\1.559 & 0.00002106 \\1.569 & 0.0001589 \\1.571 & 0.001199 \\1.571 & 0.009048 \\1.562 & 0.00002755 \\1.570 & 0.0002079 \\1.571 & 0.001569 \\1.571 & 0.01184 \\1.564 & 0.00003604 \\1.570 & 0.0002720 \\1.571 & 0.002052 \\1.571 & 0.01548 \end{tabular}
  & 1.568, 0.003117
  & \begin{tabular}{cc}
  -0.007703 & -0.003092 \\0.001683 & -0.002927 \\0.002927 & -0.001683 \\0.003092 & 0.007703
  \end{tabular}
  & \begin{tabular}{cc}
  $ -4.662\times 10^{-3} $ & $ -9.089\times 10^{-6}$ \\$ -6.179\times 10^{-4} $ & $ -6.858\times 10^{-5}$ \\$ -8.189\times 10^{-5} $ & $ -5.174\times 10^{-4}$ \\$ -1.085\times 10^{-5} $ & $ -3.904\times 10^{-3}$ \\$ -1.015\times 10^{-3} $ & $ -4.127\times 10^{-6}$ \\$ -1.345\times 10^{-4} $ & $ -3.114\times 10^{-5}$ \\$ -1.783\times 10^{-5} $ & $ -2.350\times 10^{-4}$ \\$ -2.363\times 10^{-6} $ & $ -1.773\times 10^{-3}$ \\$ 1.773\times 10^{-3} $ & $ 2.363\times 10^{-6}$ \\$ 2.350\times 10^{-4} $ & $ 1.783\times 10^{-5}$ \\$ 3.114\times 10^{-5} $ & $ 1.345\times 10^{-4}$ \\$ 4.127\times 10^{-6} $ & $ 1.015\times 10^{-3}$ \\$ 3.904\times 10^{-3} $ & $ 1.085\times 10^{-5}$ \\$ 5.174\times 10^{-4} $ & $ 8.189\times 10^{-5}$ \\$ 6.858\times 10^{-5} $ & $ 6.179\times 10^{-4}$ \\$ 9.089\times 10^{-6} $ & $ 4.662\times 10^{-3}$ \\
  \end{tabular} \\
\hline
\hline
  \end{tabular}
  \caption{$\theta$- and $\alpha$-angles for the demonstrative 3-site example considered in Appendix~\ref{app:AlphaExample}.  Within each cell, angles are ordered by increasing $k$ value or binary-interpreted controls on the associated quantum gate.
  }
  \label{tab:thetaalphaeg}
\end{table}

\begin{acknowledgments}
We would like to thank Stephen Jordan for illuminating discussions leading to this work.
We would also like to thank David Hertzog and the Center for Experimental Nuclear Physics and Astrophysics (CENPA)
for providing a creative environment for the development of this work.
We would like to thank David Kaplan and Aidan Murran for inspiring interactions.
NK and MJS were supported by the Institute for Nuclear Theory with DOE grant No. DE-FG02-00ER41132, and Fermi National Accelerator Laboratory
PO No. 652197.   This work is supported in part by the U.S. Department of Energy, Office of Science, Office of Advanced Scientific Computing Research (ASCR) quantum algorithm teams program, under field work proposal number ERKJ333.
NK was supported in part by a Microsoft Research PhD Fellowship.  We have made extensive use of Wolfram Mathematica~\cite{Mathematica} and
the quantum circuits appearing in this paper were typeset using the latex package Qcircuit, originally developed by Bryan Eastin and Steven Flammia.
\end{acknowledgments}

\bibliography{egbib}
\end{document}